\documentclass[useAMS,usenatbib]{mn2e}

\usepackage{graphicx}
\usepackage{times}
\usepackage{natbib}

\newcommand{\gtrsim}{\ga}
\newcommand{\lesssim}{\la}

\newcommand{\sbphot}{ph s$^{-1}$ cm$^{-2}$ deg$^{-2}$}
\newcommand{\sbphotsr}{ph s$^{-1}$ cm$^{-2}$ sr$^{-1}$}
\newcommand{\sbunits}{erg s$^{-1}$ cm$^{-2}$ deg$^{-2}$}

\newcommand{\lcdm}{$\Lambda$CDM}
\newcommand{\omegam}{$\Omega_{\rm m}$}
\newcommand{\omegal}{$\Omega_{\Lambda}$}
\newcommand{\omegab}{$\Omega_{\rm b}$}

\newcommand{\gadgettwo}{\textsc{gadget-2}}
\newcommand{\gadgetthree}{\textsc{gadget-3}}
\newcommand{\msun}{M$_\odot$}
\newcommand{\zsun}{Z$_\odot$}
\newcommand{\hmone}{$\,h^{-1}$}

\newcommand{\wmapfive}{{\it WMAP-5}}

\newcommand{\rosat}{{\it ROSAT}}

\newcommand{\xmm}{{\it XMM-Newton}}
\newcommand{\chandra}{{\it Chandra}}
\newcommand{\vi}{{\sc vi}}
\newcommand{\vii}{{\sc vii}}
\newcommand{\viii}{{\sc viii}}
\newcommand{\runone}{W$_{75,512}$}
\newcommand{\runtwo}{W$_{37,400}$}
\newcommand{\acf}{{\it AcF}}

\def\aap{A\&A}
\def\apj{ApJ}

\def\apjl{ApJ}
\def\mnras{MNRAS}
\def\araa{ARA\&A}

\def\nat{Nat}
\def\aaps{A\&A Supp.}
\def\apjs{ApJS}
\def\pasj{PASJ}
\def\gca{GeCoA}

\title[The effect of feedback on WHIM emission]
{The effect of feedback on the emission properties of the Warm-Hot Intergalactic Medium}
\author[M. Roncarelli et al.]
{M.~Roncarelli$^1$, 
N.~Cappelluti$^{2,3}$, 
S.~Borgani$^{4,5,6}$, 
E.~Branchini$^{7,8,9}$ and
\newauthor
L.~Moscardini$^{1,2,10}$
 \\~\\ 
$^1$Universit\`a di Bologna, Dipartimento di Astronomia, via Ranzani 1, I-40127 Bologna, Italy \\
$^2$INAF-Osservatorio Astronomico di Bologna, Via Ranzani 1, 40127 Bologna, Italy \\
$^3$University of Maryland, Baltimore County, 1000 Hilltop Circle, Baltimore, MD 21250, USA \\
$^4$Universit\`a di Trieste, Dipartimento di Fisica, Sezione di Astronomia, Via Tiepolo 11, 
     I-34143 Trieste, Italy \\
$^5$INAF-Osservatorio Astronomico di Trieste, Via Tiepolo 11, I-34143 Trieste, Italy \\
$^6$INFN, Sezione di Trieste, Via Valerio 2, I-34127 Trieste, Italy \\
$^7$Universit\`a degli Studi ``Roma Tre'', Dipartimento di Fisica ``E. Amaldi'', via della Vasca Navale 84, I-00146, Roma, Italy  \\
$^8$INFN, Sezione di ``Roma Tre'', via della Vasca Navale 84, I-00146, Roma, Italy \\ 
$^9$INAF, Osservatorio Astronomico di Brera, via Brera 28, I-20121, Milano, Italy \\
$^{10}$INFN, Sezione di Bologna, viale Berti Pichat 6/2, I-40127 Bologna, Italy \\
}
\begin{document}

\pagerange{\pageref{firstpage}--\pageref{lastpage}} \pubyear{2012}

\maketitle

\label{firstpage}

\begin{abstract}
At present, 30--40 per cent of the baryons in the local Universe is still 
undetected. According to theoretical predictions, this gas should reside in filaments filling the 
large-scale structure (LSS) in the form of a Warm-Hot Intergalactic Medium (WHIM), at temperatures 
of $10^5 - 10^7$~K, thus emitting in the soft X-ray energies via free-free interaction and line 
emission from heavy elements.
In this work we characterize the properties of the X-ray emission of the WHIM, and the LSS in 
general, focusing on the influence of different physical mechanisms, namely galactic winds (GWs), 
black-hole feedback and star-formation, and providing estimates of possible 
observational constraints. To this purpose we use a set of cosmological hydrodynamical simulations 
that include a self-consistent treatment of star-formation and chemical 
enrichment of the intergalactic medium, that allows us to follow the 
evolution of different metal species. We construct a set of simulated light-cones 
to make predictions of the emission in the 0.3--10 keV energy range.
We obtain that GWs increase by a factor of 2 the emission of both galaxy clusters and WHIM. The 
amount of oxygen at average temperature and, consequently, the amount of expected bright O\vii\ and 
O\viii\ lines is increased by a factor of 3 due to GWs and by 20 per cent when assuming a top-heavy 
IMF. 
We compare our results with current observational constraints and find that the emission from 
faint groups and WHIM should account from half to all of the unresolved X-ray background in the 
1--2 keV band.
\end{abstract}



\section{Introduction} \label{sec:intro}

The amount of baryonic matter present in the Universe is nowadays measured with high accuracy in 
the framework of the standard cosmological $\Lambda$-cold dark matter (\lcdm) scenario.
According to the most recent CMB observations \citep{komatsu11} the 
percentage of baryonic density with respect to the critical one is $100 \, \Omega_b = 4.49 \pm 
0.28$, a figure which is in excellent agreement with the estimates coming from the primordial 
nucleosynthesis of the heavy elements \citep{kirkman03} and with the baryon budget at high redshift 
estimated from Ly$\alpha$ absorption systems \citep{weinberg97,rauch98}, thus representing one 
of the most important result of 
modern cosmology. Even if they constitute a minor component in the matter-energy budget, baryons 
are indeed crucial because, since they interact with the electromagnetic field, they are 
responsible for all the radiative phenomena, thus providing us the only probe to unveil also the dark 
components.

On the other side baryons associated to Ly$\alpha$ absorbers at the present epoch account for only 
$\sim 30$ per cent 
of the cosmic mean \citep{penton04,bregman07}. Indeed, when the cosmic structures form, the baryon 
census becomes much more challenging and it requires observations at different wavelengths. 
Despite the great improvements in our observational capabilities in the last decade, when 
accounting for all the matter observed from the optical band (stars) to the X-rays (clusters) 
at $z \lesssim 2$ the amount of observed gas in the Universe is much lower than expected. More 
precisely, about 30--40 per cent of the gas is still out of reach from current observational 
instruments \citep[see][for a review]{nicastro05a}: this issue is currently referred to as the 
``missing baryons'' problem.

The first attempts to address this matter from the theoretical point view come from hydrodynamical 
simulations \citep{cen99,dave01} that showed that when galaxy clusters form, a significant amount 
of the whole intergalactic medium (IGM) remains out of the virialised halos creating a network of 
filaments, the so-called cosmic web \citep{bond96}, that provide a continuous flow of accreting 
material to the clusters themselves, located at the knots of the structure. The gas in 
these filaments remains at moderate overdensities ($\delta \approx 10 - 10^3$) and 
is shock-heated up to temperatures of the order of $10^5 - 10^7$ 
K, and is usually addressed as the Warm-Hot Intergalactic Medium (WHIM). Given its thermodynamical 
conditions the hydrogen and helium are fully ionized and emit via free-free interaction with the 
electrons. This process together with the line-emission from partly ionized metals is expected to 
contribute to the X-ray background (XRB) in the form of a soft diffuse component.

Analysing the current observational constraints, the 
unresolved emission in the \chandra\ Deep Fields (CDFs) observed by \cite{hickox07} in the 0.5--1 
keV band is consistent with the expectations of 
the diffuse gas obtained from numerical simulations \citep[see, e.g.,][]{roncarelli06}. Moreover, 
\cite{galeazzi09} 
detected the signature of a possible WHIM emission in the correlation function of the diffuse 
emission signal in the 0.4--0.6 keV band obtained from several \xmm\ observations.
On the other side, there have also been claims of detection of WHIM emission associated to local 
overdensities of the LSS. \cite{werner08} observed at 5$\sigma$ a continuum emission from a 
filament extending between the Abell 222 and Abell 225 clusters. Similarly, \rosat\ 
observations allowed the discovery of a soft diffuse emission consistent with $T \approx 10^6$ K 
gas associated with a galaxy group at $z\sim 0.45$ \citep{zappacosta02} and with the Sculptor 
supercluster \citep{zappacosta05}.

Probably the most important point in WHIM observation strategies is the fact that this medium is 
believed to contain metals whose emission and absorption lines provide an effective way to 
disentangle its signal from background and foreground objects. To this purpose, the best targets are 
C, O, Ne and possibly also N, Mg and Fe. The brightest features are expected from K and L-shell 
transitions of H-like, He-like and Li-like oxygen ions, namely the O\viii\ $1s-2p$ doublet 
($E=$654 eV), the O\vii\ $1s-2p$ resonance line (574 eV) and the O\vi\ doublet (12 eV), respectively.
This method already provided several claimed detections, all regarding absorption lines. 
\cite{danforth05,danforth08} detected 83 O\vi\ absorption systems in the UV spectra of 43 AGNs. 
However, since 
these absorbers are associated to Ly$\alpha$ counterparts it is not clear how much of this gas can 
be attributed to the missing part \citep{tripp08}. At X-ray energies \cite{nicastro05b} detected 
several O\vii\ absorption features in the spectra of the Mrk 421 blazar, but the statistical 
significance of their measurements has been questioned \citep{kaastra06,rasmussen07,yao11}. Several 
other 
detections with both \chandra\ \citep{fang02,mathur03,zappacosta10} and \xmm\ \citep{fujimoto04,
takei07,buote09,fang10} have been claimed but their statistical significance is low and they mainly 
correspond to the densest part of the IGM (galaxy clusters or groups), indicating that these 
measurements are at the limit of current observational capabilities.

Up to now the detection of emission lines has proven to be even more difficult. 
However, detecting the WHIM in emission can be much more rewarding since it has the potential to 
map large structures and does not depend on the presence of background sources, thus potentially 
providing a global picture of WHIM thermo- and chemo-dynamics, which is by now far from being 
reached.

In this scenario a big effort has been put into the modelisation of the WHIM which becomes crucial 
in the perspective of future X-ray missions in particular for what concerns the treatment of 
radiative cooling, stellar and galactic feedback, reionization and, most importantly, chemical 
enrichment. \cite{cen06a} and \cite{cen06b} included for the first time the enrichment from Type II 
supernovae (SN-II) of different metal species in their Eulerian cosmological simulations. Their 
work showed how the effect of different galactic wind models can change both the thermodynamics and 
the abundance outside clusters. More recently \cite{bertone10} used a similar approach with a 
set of several large-scale simulations performed with the \gadgetthree\ code, including also Type 
Ia supernovae (SN-Ia) and asymptotic giant branch (AGB) stars to study specifically the properties 
of metal line emission.

In this work we use the set of cosmological hydrodynamical simulations of \cite{tornatore10} to 
study and characterize the emission of the WHIM, with particular emphasis on the of impact 
different enrichment schemes. Starting from the outputs of these simulations, we adopt a light-cone 
reconstruction approach in order to provide a more direct comparison with current and future 
observational constraints.

This paper is organized as follows. In the next section we will describe the simulation set of 
\cite{tornatore10} which constitutes the basis of our study. In Section~\ref{sec:lcones} we will 
explain our method to create the mock light-cones and spectra from the outputs of the 
hydrodynamical simulations. Section~\ref{sec:emission} is devoted to the analysis of the global 
properties of the IGM and WHIM emission, while Section~\ref{sec:oxygen} focuses on our predictions 
for the oxygen line statistics. We summarize our results and draw our conclusions in 
Section~\ref{sec:concl}. Unless stated otherwise, when referring to solar metal abundances and 
relative element yields we will assume the results reported in \cite{asplund09}.


\section{The hydrodynamical simulations} \label{sec:sims}

In this section we review the simulation set of \cite{tornatore10}. We refer the 
reader to their original work for further details. The set consists of 8 cosmological 
hydrodynamical simulations performed with the TreePM-smoothed particle hydrodynamics (SPH) code 
\gadgettwo\ \citep{springel05} in a version that includes the implementation of the chemical 
enrichment of \cite{tornatore07}. A summary of the main characteristics of the different runs is 
shown in Table \ref{tab:sims}.

\begin{table*}
\begin{center}
\caption{
Characteristics of the 8 simulations object of our study. Column 1: run name. Column 2: comoving 
length of the box ($h^{-1}$ Mpc). Column 3: number of particles. Column 4: number of available 
snapshots in the interval $0 < z < 1.5$. Column 5: stellar IMF assumed. Column 6: feedback scheme 
(see more details in the text).}
\begin{tabular}{lccccc}
\hline
\hline
Run name     & $L_{\rm box}$ & $N_{\rm dm}, N_{\rm gas}$ & Snapshots &        IMF      &     Feedback model    \\
\hline
W (reference)&      37.5     &    256$^3$   &     25    &      Kroupa     &          Winds        \\              
W$_{75,512}$ &      75.0     &    512$^3$   &      6    &      Kroupa     &          Winds        \\              
W$_{37,400}$ &      37.5     &    400$^3$   &     16    &      Kroupa     &          Winds        \\              
Way          &      37.5     &    256$^3$   &     25    &  Arimoto-Yoshii &          Winds        \\              
Ws           &      37.5     &    256$^3$   &     25    &     Salpeter    &          Winds        \\              
NW           &      37.5     &    256$^3$   &     25    &      Kroupa     &        No winds       \\              
CW           &      37.5     &    256$^3$   &     25    &      Kroupa     &     Coupled winds     \\              
BH           &      37.5     &    256$^3$   &     25    &      Kroupa     & BH-feedback, no winds \\              
\hline
\hline
\label{tab:sims}
\end{tabular}
\end{center}
\end{table*}

For all the runs the cosmology assumed is a flat standard \lcdm\ model consistent with the 
\wmapfive\ results \citep{komatsu09}, namely with \omegal=0.76, \omegam=0.24 and \omegab=0.0413 for 
the matter-energy density parameter of the cosmological constant, total matter (dark and baryonic) 
and baryonic matter alone, respectively. The other cosmological parameters are a primordial power 
spectrum of the cold dark matter density perturbations with slope $n_s=0.95$ and normalization 
corresponding to 
$\sigma_8$=0.8 and a Hubble constant $H_0=$100$\,h$ km s$^{-1}$ Mpc$^{-1}$, with $h=0.73$. The 
gravitational softening is set to $\varepsilon=7.5 h^{-1}$ kpc in comoving units at $z>2$ while at lower 
redshifts it is set to $\varepsilon=2.5 h^{-1}$ physical kpc.

We take the W run (see Table \ref{tab:sims}) as a reference model for comparison. This simulation 
follows the evolution of a box of $L=$37.5$\,h^{-1}$ comoving Mpc per side, with 2$\times$256$^3$ 
particles for dark matter and gas: this results in a mass resolution of $m_{\rm dm}=1.9 \times 
10^8 h^{-1}$ \msun\ and $m_{\rm gas}=3.6 \times 10^7 h^{-1}$ \msun, respectively. The choice of 
these parameters is the result of a compromise between the need to sample the perturbations on 
large scales and to achieve the resolution necessary to describe the dense environments where star 
formation and feedback take place. This 
run produced 43 outputs logarithmically equispaced in redshift from $z=4$ to $z=0$. Since 
we are interested mainly in the WHIM emission and we do not expect significant contributions from 
high redshift gas, for the purpose of our work we considered only the 25 snapshots in the interval 
$0<z<1.5$.

The effect of a spatially uniform, redshift dependent UV background due to quasar radiation as in 
\cite{haardt96} is considered \citep[with helium heating rates enhanced by a factor of 3, see the 
discussion in][]{tornatore10}. This process is important when considering the evolution of the 
amount of matter in the WHIM phase \citep[see, e.g.,][]{dave01}. Star formation is modeled 
following the multiphase model of \cite{springel03}: in this prescription the SPH particles 
exceeding the density of $n_{\rm H}$=0.1 cm$^{-3}$ are considered to host a hot phase and a cold 
one in 
pressure equilibrium, with the latter assumed to host the star-forming regions. Every two-phase SPH 
particle spawns up to three generation of collisionless star particles in a stochastic way, each 
of them having about one third of the mass of the original gas one.

During this process the SPH particles are enriched following the prescription of \cite{tornatore07} 
that considers the formation of SN-II, SN-Ia together with low and intermediate 
mass stars in the AGB phase. The corresponding stellar yields are taken from \cite{woosley95}, 
\cite{thielemann03} and \cite{vandenhoek97} for the SN-II, SN-Ia and 
AGB, respectively. This code allows to follow independently the evolution of H, He and of 6 metal 
species, namely C, O, Mg, Si, S and Fe for every SPH particle. The formation of all the other 
elements is also computed in every gas particle though only globally, storing their total mass in a 
single variable in order to save memory and computation time.  The IMF adopted is the one proposed 
by \cite{kroupa01}, with a multi-slope function of the form $\varphi(m) \propto m^{-y}$, with the 
index $y$ varying in the range 0.3--1.7 from low to high masses.

Radiative cooling due to free-free interaction of the electrons and ions is computed using a 
primordial mix of H and He. In addition to that, cooling from heavy elements is followed adopting 
the cooling rates of \cite{sutherland93}: in this scheme the global metallicity of every SPH 
particle, and not the individual elements, is considered assuming solar yields. However, as we will 
explain in detail in Section \ref{sub:spectra}, we stress 
that for what concerns the results presented in this paper our mock emission spectra are computed 
considering all the elements provided by the \cite{tornatore07} code separately. We 
point out also that this cooling scheme assumes implicitly that the IGM is in collisional 
ionization equilibrium: although this is not completely true, this can be considered as a safe 
approximation of the gas ionization state in the densest regions of the WHIM that dominate the 
emission.

Galactic outflows are modeled following \cite{springel03}. Winds are assumed to blow from the 
star-forming regions with a mass loading rate $\dot{M}_w = \eta \dot{M}_\star$, where 
$\dot{M}_\star$ is the star-formation rate of each multi-phase particle and the efficiency $\eta$ 
is a free parameter (set to 2). SPH star-forming particles are then stochastically selected to 
become wind particles, with a 
velocity of $v_{\rm w}$=500 km/s and are decoupled from the others allowing them to travel freely 
for 10$\, h^{-1}$ kpc. This allows the wind particles not to get stuck inside the dense 
star-forming regions.

In order to test the impact of the different parameters and assumptions of our reference W model, 
we analyze also the results of other 7 runs of \cite{tornatore10}. It is important to note that all 
the simulations were run assuming the same initial conditions so that they actually follow the 
evolution of the same volume and matter (with the exception of \runone\ that follows a larger 
volume) thus allowing a direct comparison between each other. In the following subsections we will 
describe these runs focusing on their differences with respect to the reference model.


\subsection{Simulation parameters}
The \runone\ and \runtwo\ runs assume the same reference model but 
with different volume size and resolution of the particles. These simulations can be used to assess 
how the limited volume and finite mass resolution affect our results.

The \runone\ run has a box size of $L=$75$\,h^{-1}$ comoving Mpc. The number of particles is 
augmented to 2$\times$512$^3$ in order to maintain the same resolution. On the other side the 
\runtwo\ run follows the same volume but with higher resolution. The number of particles is 
increased to 2$\times$400$^3$, achieving a resolution of $m_{\rm dm}=5.0 \times 10^7 h^{-1}$ \msun\ 
and $m_{\rm gas}=9.4 \times 10^6 \, h^{-1}$ \msun. The softening is also reduced to 
$\varepsilon=4.8 \, h^{-1}$ comoving kpc at $z>2$ and $\varepsilon=1.6 \, h^{-1}$ physical kpc at 
$z<2$.

The number of snapshots available for these runs is reduced. In the interval $0<z<1.5$ we have 6 
snapshots for \runone\ and 16 for \runtwo. However, our scheme for the reconstruction 
of the observed light-cone (see Section \ref{sub:conevol}) allows us to minimize the impact of 
these differences.


\subsection{Initial mass functions}
The Way and Ws runs differ from the reference model as the IMF adopted for these two runs are from 
\cite{arimoto87} and \cite{salpeter55}, respectively. Changing the IMF implies a modification of the 
proportion of the SN types which reflects into different element ratios: top-heavier models produce 
a relatively higher number of SN-II and therefore are expected to provide more oxygen with respect 
to iron and carbon.

Indeed, the Way run shows a global value of [C/O]=--0.3 for the WHIM at $z=0$ while the  W and Ws 
runs have values close to --0.15. All the three models show a similar behavior 
of the redshift evolution with [C/O] ratios growing with time due to the increasing contribution of 
SN-Ia \citep[see the corresponding discussion in][]{tornatore10}.


\subsection{Feedback models}
The NW, CW and BH runs adopt a modified feedback scheme with respect to our reference W run. In the 
NW run galactic winds are simply turned off, while the CW simulation keeps wind particles 
hydrodynamically coupled with the other ones \citep[more details about the effect of coupled winds 
on star-forming systems can be found in][]{dallavecchia08}.

The BH run assumes no winds but includes the energy feedback associated to supermassive black-holes 
(SMBH) following \cite{springel05} and \cite{dimatteo05}. In brief, SMBHs are computed as sink 
particles hosted by haloes of mass higher than $M_{\rm th} = 10^{10}$\hmone\msun\ that grow with a 
rate of
\begin{equation}
\dot{M}_{\rm BH} = {\rm min}(\dot{M}_{\rm B},\dot{M}_{\rm Edd}) \ ,
\end{equation}
where $\dot{M}_{\rm B}$ and $\dot{M}_{\rm Edd}$ are the Bondi-Hoyle-Lyttleton accretion rate 
\citep[see, e.g.,][]{bondi52} and the Eddington rate, respectively. The energy rate injected into 
the surrounding gas is computed as $\dot{E}_{\rm feed} = \epsilon_{\rm r}\epsilon_{\rm f} 
\dot{M}_{\rm BH}^2$, being $\epsilon_{\rm r}$ and $\epsilon_{\rm f}$ the two free parameters 
representing the radiative efficiency and the fraction of energy coupled to the gas, respectively. 
These values are set to $\epsilon_{\rm r}=0.1$ and $\epsilon_{\rm f}=0.05$ with which 
\cite{dimatteo05} were able to reproduce the observed relation between BH mass and velocity 
dispersion of the bulge.

We refer the reader to the detailed discussion in \cite{tornatore10} about the effects of these 
models on the properties of the IGM. As a general remark, here we point out that NW and BH models 
show a stronger star formation rate (SFR) at higher redshift 
with respect to our reference run. Starting at $z\sim 3$ the BH feedback causes a heavy suppression 
of star-formation, getting closer to the W simulations, while for the NW the SFR stays higher 
even at later epochs. It is important to highlight also that BH feedback is able to affect 
significantly the properties of collapsed objects causing the IGM to be more diffuse at all scales.


\section{The simulated light-cones} \label{sec:lcones}

Extracting observables from the outputs of LSS simulations requires a non-trivial post-processing 
analysis: indeed, one needs to reconstruct all the volume of the past light-cone as seen by an 
observer located at $z=0$ in order to account for all the matter that contributes to the integrated 
signal up to a given redshift.

For the purpose of estimating the properties of the WHIM emission, we 
decide to consider the interval $0<z<1.5$: this is enough to account for most of the expected 
signal \citep[see][]{roncarelli06,takei11}. Then we compute the X-ray emission spectra 
associated to the gas elements present in the light-cone volume, we integrate them along 
the line-of-sight and, finally, we produce mock 3D-maps in  which every pixel contains the spectrum 
as seen by a telescope.

In this Section we describe in detail the different steps of this procedure.


\subsection{Reconstructing the light-cone volume} \label{sub:conevol}
The method described here is an improvement of the one adopted in \cite{roncarelli06} and 
\cite{ursino10} for the X-ray emission of the LSS \citep[see also similar applications for 
estimates of the Sunyaev–Zel'dovich effect in][]{dasilva00,dasilva01,white02,zhang02,roncarelli07}, since it 
includes a more general randomization scheme and a more precise procedure for the redshift sampling 
\citep[as in][]{roncarelli10}.

We reconstruct the light-cone volume by replicating the outputs of the simulations along the line of 
sight up to the redshift limit of $z=1.5$. With the cosmological parameters assumed here 
(see Section \ref{sec:sims}) this corresponds to a comoving distance of $d_{\rm c,lim}=3206$\hmone 
Mpc, which is 86 times the length of our simulation boxes (43 for the \runone\ run). 
When considering 
large distances one needs also to take into account the evolution of structure formation by using 
the different simulation snapshots. To this purpose for a given distance from the observer we chose 
the output that better approximates the age of the Universe at that 
epoch. Since in our case we have more cubes to stack than available outputs, most of the cubes will 
correspond to the same snapshot than their neighbor(s). When a snapshot change intervenes the cube 
is split into two slices at the corresponding distance.

In order to avoid the repetition of the same structures in our maps, every cube undergoes a 
series of randomization procedures. First we give to every axis a 50 per cent probability of being 
reflected, then we chose randomly the axis order. Once this is done, we exploit the periodic 
boundary conditions of our simulations to select a position inside the cube which is then put in 
its center. Finally we introduce a rotation of a random angle along the line of sight only if the 
distance from the observer is small enough to ensure that the projected cube encloses all the map 
surface. As in our previous works, the slices corresponding to different snapshots but belonging to 
the same cube undergo the same randomization process to preserve power on large scales.


\subsection{Computing the X-ray spectra and maps} \label{sub:spectra}
For every SPH particle that falls inside the light-cone we compute its spectrum deriving it from 
its physical variables. We adopt the \textsc{apec} emission model \citep[][version 2.0\footnote{We 
are aware of the error in the calculation of some ions present in this version of the code. We 
checked that its impact on our results is negligible.}]{smith01} which assumes that the emitting 
gas is an optically thin plasma in collisional ionization equilibrium. We create a series 
of emission tables by using the \textsc{vapec} implementation on \textsc{xspec} (version 12.6.0) 
that allows to separate the contribution of the different elements. More in detail, for every 
element considered in our simulations (H, He, C, O, Mg, Si, S, Fe and the sum of 
all the others\footnote{The relative abundances of the elements not considered individually are 
computed from \cite{asplund09}.}, as 
described in Section \ref{sec:sims}) we extract three-dimensional tables $s_{\rm el}(z,T,E)$ with 
fixed metallicity and normalization in which two variables are the redshift and temperature and the 
third one is the energy channel. 
Then the flux as a function of energy $E$ is computed for the $i$-th SPH particle as
\begin{equation}\label{eq:spec}
F_i(E) = \frac{n_{{\rm e},i} \, n_{{\rm H},i} \, V_i} {4\pi d_{{\rm c},i}^2} 
       \sum_{\rm el} X_{{\rm el},i} \, s_{\rm el}(z_i,T_i,E) \ ,
\end{equation}
where $n_{{\rm e},i}$ and $n_{{\rm H},i}$ are the electron and hydrogen number density, 
respectively, for each particle, $V_i$ is its physical volume and $d_{{\rm c},i}$ is its comoving 
distance from the observer. The redshift $z_i$ and the temperature $T_i$ of the SPH particle are 
used to interpolate the value of $s_{\rm el}(z_i,T_i,E)$ from the tables. The abundance of each 
element is defined as
\begin{equation}\label{eq:abund}
X_{{\rm el},i}\equiv A_{\rm el} \, \frac{n_{{\rm el},i}}{n_{{\rm H},i}} \ ,
\end{equation}
where $A_{\rm el}$ is the element atomic mass in units of the proton mass and $n_{{\rm el},i}$ is 
the number density of the element provided by the code\footnote{In the formula of 
eq.~(\ref{eq:spec}) the abundance is generalized also for hydrogen for which, of course, 
$X_{\rm H}\equiv1$.}.

Finally, in order to obtain the surface brightness the flux of every particle computed from 
eq.~(\ref{eq:spec}) is distributed in the map pixels by integrating analytically the SPH 
smoothing kernel \citep[the details of the procedure are explained in][]{ursino10}.


\subsection{The light-cone set}\label{sub:coneset}
The method described above allows us to compute a set of 3-d mock spectral maps (i.e. the 
surface brightness $SB(\vec{\theta},E)$, as a function of the position in the sky $\vec{\theta}$ 
and energy $E$) from which we extract our results. 

\begin{table*}
\begin{center}
\caption{
Summary of the sets of 3-d spectral maps created for our studies. In all cases the redshift 
interval is $0<z<1.5$. Column 1: total number of mock maps (the 
number of simulation runs times the number of independent light-cones). Column 2: gas phase, IGM 
corresponds to all of the gas present in the simulation outputs, WHIM corresponds 
to the gas with temperature in the range $10^5-10^7$ K and density contrast $\delta<1000$. Column 
3: chemical elements considered. Column 4: angular size if the maps. Column 5: angular 
resolution. Column 6: energy range of the spectrum. Column 7: energy resolution for the 
$E<1$ keV bins.}
\begin{tabular}{ccccccc}
\hline
\hline
$N$                   &   Gas phase  & Elements & Map size & Ang. resolution &  Energy range &  En. res. ($E<$1 keV) \\
(sims $\times$ cones) &              &          &   (deg)  & (arcsec)  &  (keV)        &        (eV)       \\
\hline
$8 \times 20$         &   IGM (all)  & All & 0.5 &  3.52 & 0.3--10 & 50 \\
$8 \times 20$         &   WHIM       & All & 0.5 &  3.52 & 0.3--10 & 50 \\
$8 \times 20$         &   IGM (all)  &  O  & 0.5 & 56.25 & 0.3--1  &  1 \\
$8 \times 20$         &   WHIM       &  O  & 0.5 & 56.25 & 0.3--1  &  1 \\
\hline
\hline
\label{tab:cones}
\end{tabular}
\end{center}
\end{table*}

To assess the statistical robustness of our results we construct 20 different light-cones by 
varying the initial numerical seeds necessary for the randomization process (see 
Section~\ref{sub:conevol}). Each of them is applied to all of the simulation runs, ensuring 
that every light-cone represents the same comoving volume (except for the \runone\ run), simulated 
with different physical prescriptions. This is important because it allows us to compare directly 
the results of our simulations without dealing with uncertainties associated to cosmic variance. 
On the whole we obtain 640 outputs with different characteristics summarized in 
Table~\ref{tab:cones}.

In our analysis we distinguish between the properties of the IGM as a whole and of the WHIM. Here 
we adopt the definition of WHIM as the gas with temperature in the range 10$^5$--10$^7$ K and 
density contrast $\delta < 1000$. We advise the reader that not all the authors adopt the same 
definition: in some works the WHIM is defined only with respect to the temperature cut, in 
particular this applies to \cite{roncarelli06} and to \cite{tornatore10} with which we will make 
an extensive comparison. In practice, the difference consists in the inclusion or not of the 
warm-hot dense gas that roughly corresponds to galaxy groups. As a reference, with our 
definition the WHIM accounts for 33 per cent of the total baryonic mass at $z=0$ in the W run.

Given the redshift limit of $z<1.5$, the maximum angular extension of the maps is fixed by the box 
length and corresponds to about 40 (80 for the \runone\ run) arcmin, however we chose to compute 
maps of 30 arcmin per side in order to avoid problems caused by border effects: this size roughly 
corresponds to the double of the CDFs, so that every mock map covers four times as much sky 
surface.

The first two sets of mock maps, as indicated in Table~\ref{tab:cones}, have 512 pixels per side 
which corresponds to a resolution of 3.52 arcsec. The spectra span the energy range 0.3--10 keV 
and are sampled with bins of 50 eV for energies lower than 2 keV. At higher energies we only 
compute the surface brightness of four 2 keV-wide bands. The results of these two sets will be 
compared mainly with \chandra\ and \xmm\ data in the following section.

Our method of spectra computation, as described by eq.~(\ref{eq:spec}), allows us to separate the 
contribution of the different chemical elements. Taking advantage of this, we can artificially 
``switch off'' the emission of all other elements in order to focus on the properties of oxygen 
only. To this purpose we compute two other sets of mock maps that include only the signal from 
this element.
Since our goal is to study line emission, we increase the spectral resolution to 1 eV, while 
reducing the energy range to an upper limit of 1 keV and the angular resolution to $\sim 1$ arcmin. 
We produce other two sets (for the IGM and the WHIM) with this method and use it to predict oxygen 
line counts (see Section~\ref{sec:oxygen}).


\section{Properties of the X-ray emission} \label{sec:emission}

\begin{figure*}
\includegraphics[width=1.0\textwidth]{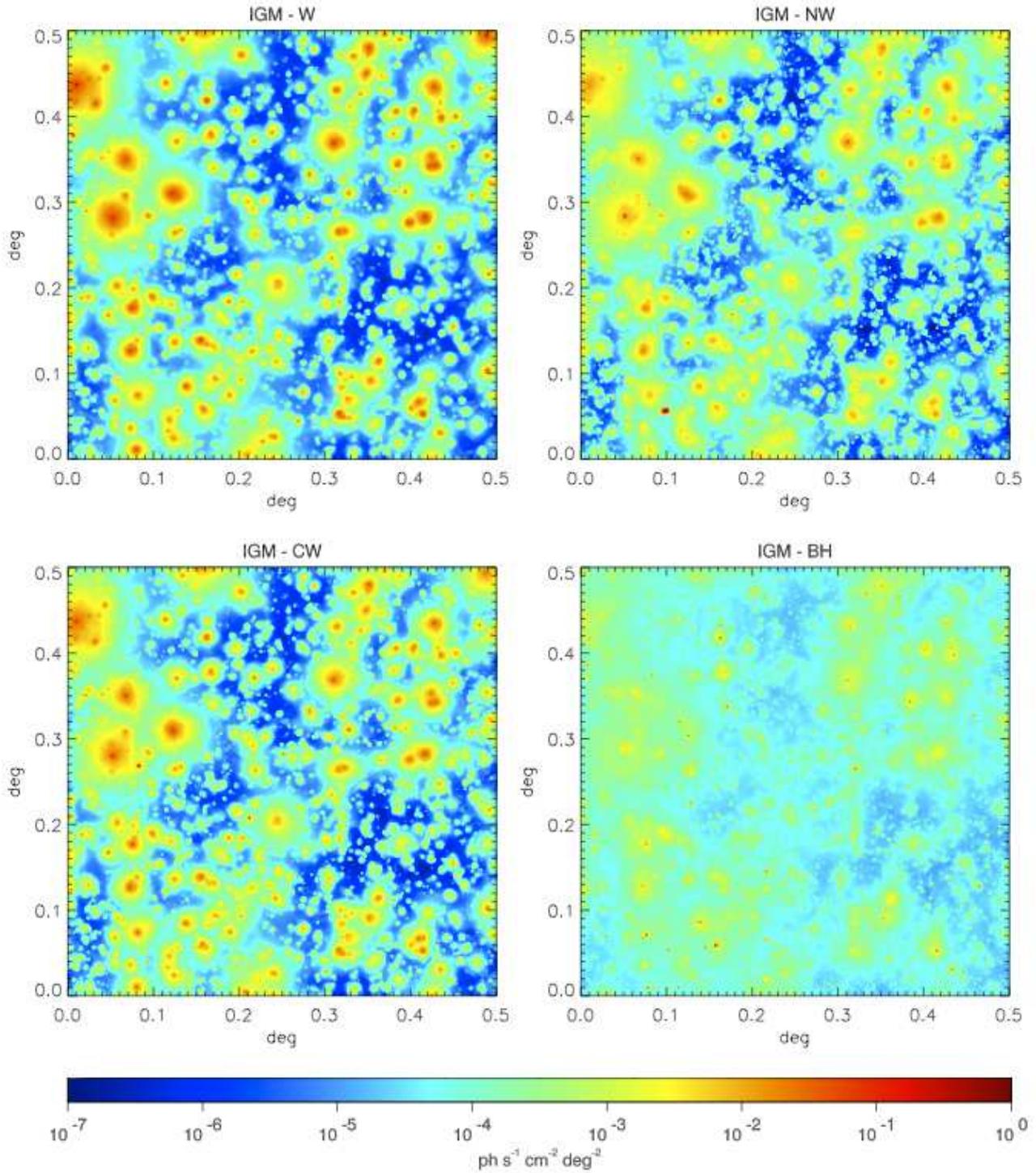}
\caption{Surface brightness of the IGM (i.e. all the gas) in the 0.3--0.8 keV band. The maps are 
0.5 deg per side and represent the same light cone simulated with different feedback schemes: 
galactic winds (W model, top left), no winds (NW model, top right), coupled winds (CW model, bottom 
left) and black-holes feedback (BH model, bottom right).}
\label{fig:maps_tot}
\end{figure*}

We show in Fig.~\ref{fig:maps_tot} the expected IGM surface brightness in the 0.3--0.8 keV band for 
a single light-cone with 4 different feedback schemes. As said in the previous section, with 
our volume reconstruction procedure we ensure that our maps reproduce exactly the same structures 
(i.e. the same objects in the same positions), with no difference associated to cosmic variance.

\begin{figure*}
\includegraphics[width=1.0\textwidth]{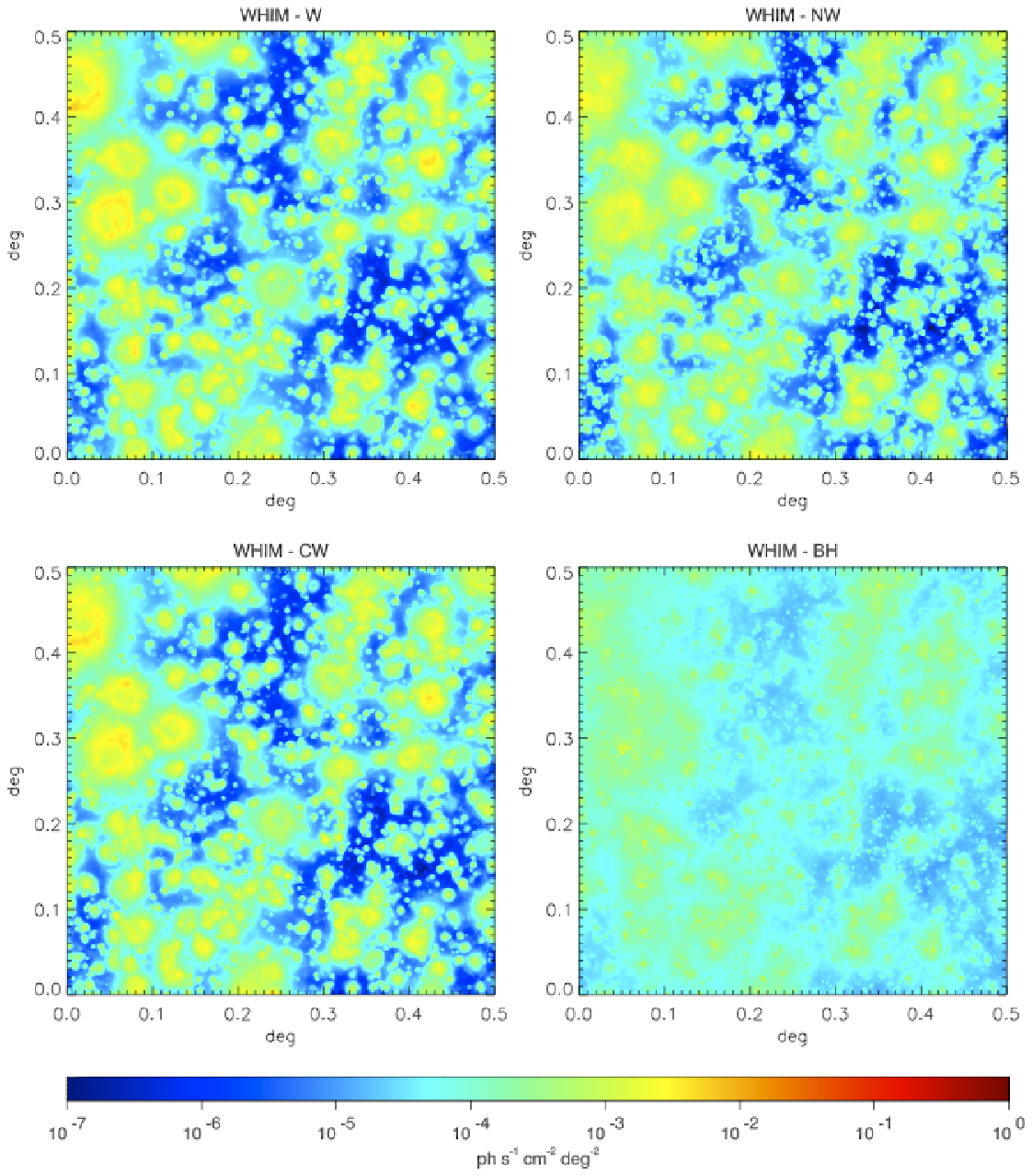}
\caption{The same as Fig.~\ref{fig:maps_tot} but for the WHIM (i.e. gas with $10^5$ K$<T<10^7$ K 
and $\delta < 1000$).
}
\vspace{1.0cm}
\label{fig:maps_dif}
\end{figure*}

The emission is dominated by galaxy clusters, with peaks that have similar values for 
all the models ($\sim$1 \sbphot). A significant fraction of the maps has a fainter  
surface brightness from 10$^{-3}$ down to 10$^{-7}$ \sbphot. In the W model the haloes are significantly 
more diffuse: given the relatively small size of 
our simulation volume we should consider this as representative of the effect at the galaxy group 
scale.
When turning winds off (NW model) clusters have sharper peaks, while the CW model produces an 
intermediate effect. In the BH model the cluster 
emission is significantly lower: the strong energy injection at early epochs increases 
considerably the entropy of the gas causing a suppression of gas accretion from high redshift. As 
a result smaller haloes are disrupted while for bigger ones only the central peaks are present 
\citep[see the discussion in][]{tornatore10}. On 
the contrary, with this model low surface brightness regions are slightly brighter: apart from 
cluster centers all pixels stay roughly within 2-3 orders of magnitudes in surface brightness. Due 
to the very strong effect on galaxy clusters and groups, this model has to be considered as an 
extreme and non-realistic case. It is, however, useful as a case study to understand the effect of 
this kind of feedback.

The maps of Fig.~\ref{fig:maps_dif} show the surface brightness of the same light-cones but 
considering the WHIM only: the difference with respect to the previous ones is of course the 
absence of the hot bright cluster centers. These images show that even if our definition of WHIM 
includes the density cut at $\delta<1000$, cluster and group outskirts close to the virialized 
regions are still present. However, these structures can be considered below the 
detection limit of the CDFs (see Section~\ref{sub:xrb} and Fig.~\ref{fig:mapreg}).
It is important to note that even when excluding artificially the hot gas directly from the 
simulations, no filamentary structure is clearly visible in these maps. This is due to the 
superimposition of the structures of the cosmic web along the line of sight with the effect of 
self-washing out.
Also in these maps the most important differences are in the BH model, where the WHIM appears more 
diffuse and fainter than in the other cases. The effect of enhancing the surface brightness of the 
faintest pixels has little impact on the possibility of detecting these regions since they still 
fall well below the limits of current instruments.
The images of the other models (not shown in these figures) do not differ significantly from our 
reference one (W model).

\begin{figure}
\includegraphics[width=0.30\textwidth,angle=-90.]{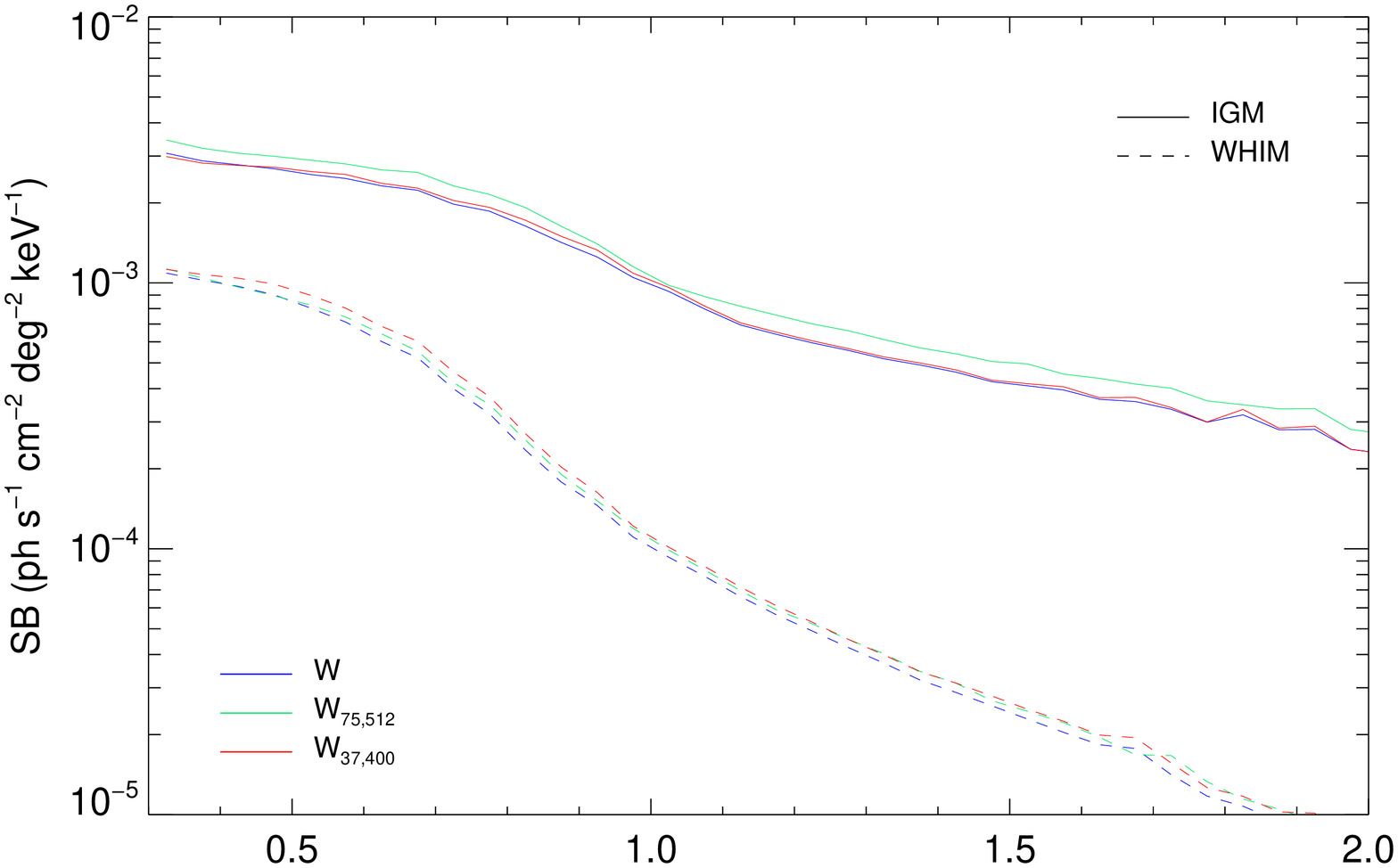} \newline
\includegraphics[width=0.30\textwidth,angle=-90.]{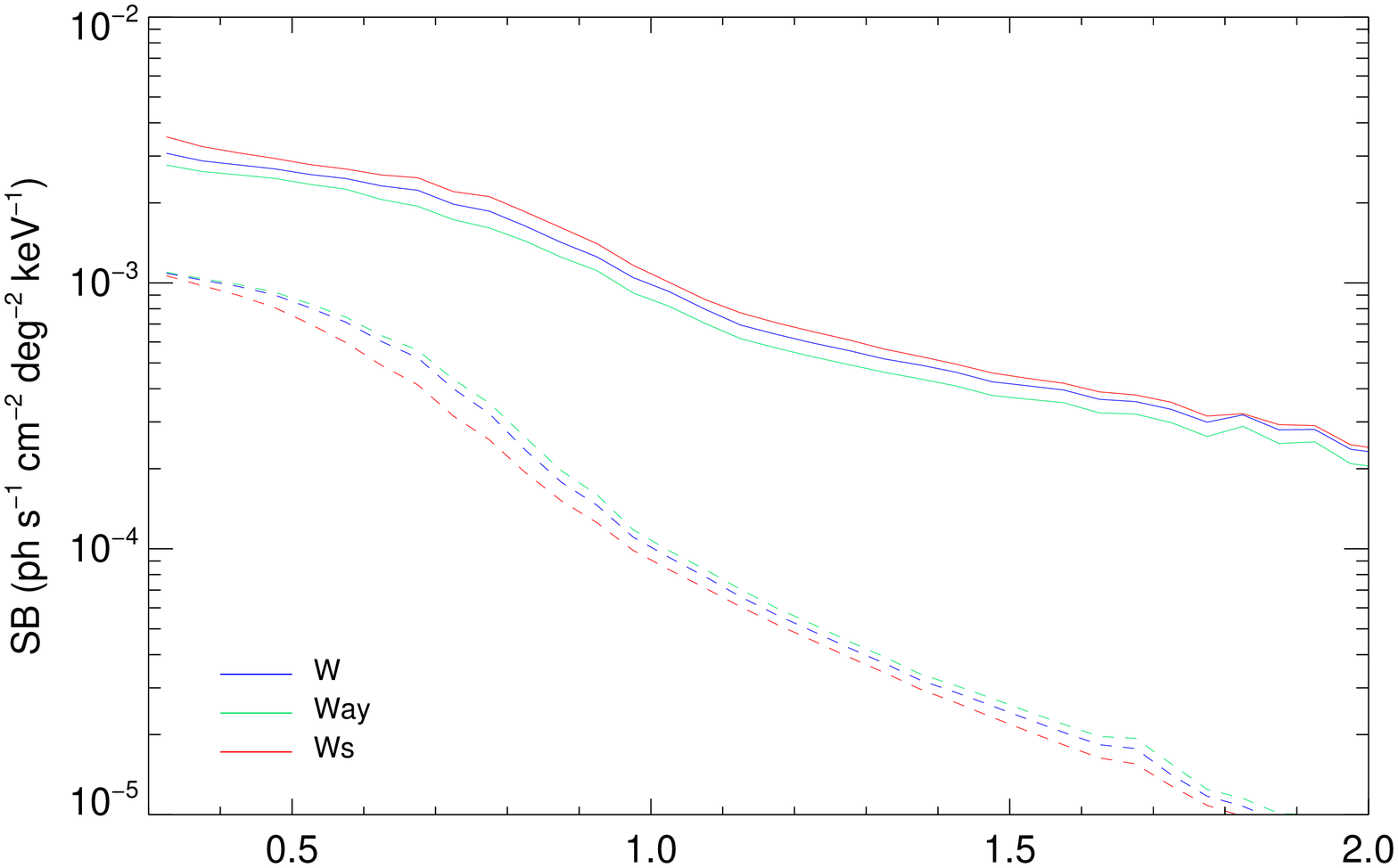} \newline
\includegraphics[width=0.30\textwidth,angle=-90.]{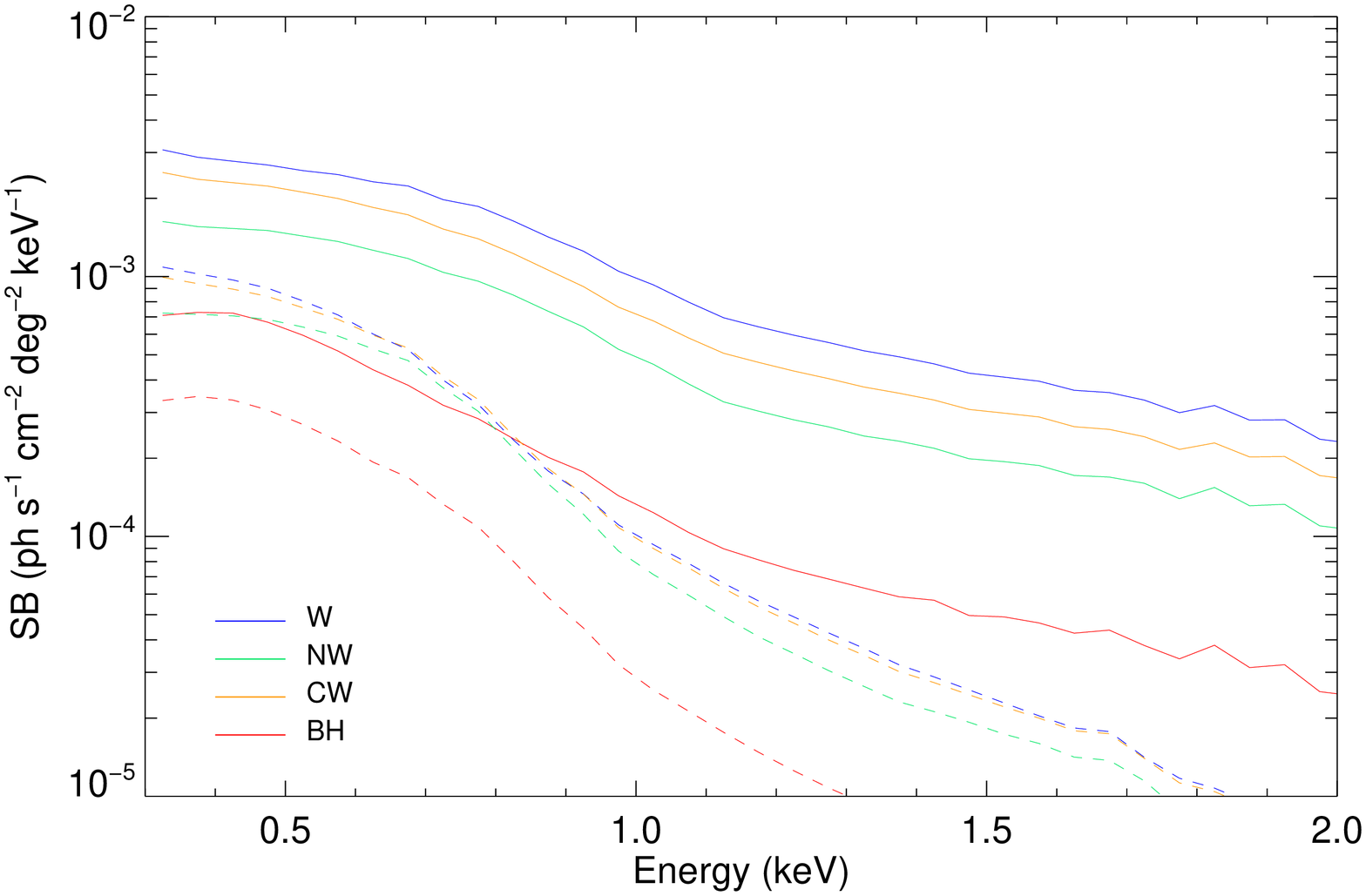}
\vspace{0.2cm}
\caption{Expected spectra of the IGM (solid lines) and WHIM (dashed lines), computed as average 
over the 20 independent light-cones. In all plots the blue 
lines correspond to our reference model (W). Top panel: comparison  with \runone\ (green) and 
\runtwo\ (red). Central panel: comparison with Way (green) and Ws 
(red). Bottom panel: comparison with NW (green), CW (orange) and BH (red).}
\label{fig:spectot}
\end{figure}

We analyse the effect of feedback on the spectral properties of the IGM and the WHIM  with the 
plots shown in Fig.~\ref{fig:spectot} and with the corresponding surface brightness in different 
bands shown in Table~\ref{tab:sb}. 
When considering a large box (\runone, top panel), the global IGM emission is about 20 
per cent higher at all energies due to the presence of more massive clusters. On the 
contrary, no significant change is produced when considering the WHIM emission, as an indication 
that our box size is large enough to sample the LSS perturbations that influence the WHIM physics. 
The opposite effect applies when increasing the resolution of our simulations (\runtwo): almost no 
difference can be seen for the IGM signal while the WHIM emission increases by $\sim$10 per cent. 
This is mainly due to the effect of resolution on metal formation/emission, as we will discuss in 
Section \ref{sec:oxygen}. On the whole we can safely conclude that the limited box size and 
resolution of our simulations have a negligible impact on the expected surface brightness of the 
WHIM.

\begin{table*}
\begin{center}
\caption{Expected surface brightness (in units of 10$^{-12}$ \sbunits) for the whole IGM (i.e. all 
the gas) and for the WHIM (i.e. gas with $T=10^5-10^7$ K and $\delta < 1000$) in different X-ray 
bands. In each line the results are shown for the 8 different simulations as the mean computed over 
20 independent light-cones (a total of 5 deg$^2$). The errors represent the standard deviations in 
fields equal to a quarter of the original maps (e.g. 225 arcmin$^2$, close to the size of 
the CDFs).}
\begin{tabular}{lcccccccccccc}
\hline
\hline
      & & \multicolumn{11}{c}{Surface brightness (10$^{-12}$ \sbunits)} \\
      & & IGM & WHIM & & IGM & WHIM & & IGM & WHIM & & IGM & WHIM \\
Model & & \multicolumn{2}{c}{(0.3--0.8) keV} & & \multicolumn{2}{c}{(0.65--1) keV} & & \multicolumn{2}{c}{(0.5--2) keV} & & \multicolumn{2}{c}{(2--10) keV} \\

\hline
W & & 1.05$\pm$1.08 & 0.29$\pm$0.10 & & 0.73$\pm$1.14 & 0.12$\pm$0.09 & & 2.12$\pm$3.84 & 0.28$\pm$0.16 & & 0.52$\pm$1.84 &$<$0.01\\
W$_{75,512}$ & & 1.20$\pm$0.90 & 0.30$\pm$0.13 & & 0.84$\pm$0.86 & 0.13$\pm$0.10 & & 2.45$\pm$2.72 & 0.30$\pm$0.18 & & 0.61$\pm$1.28 &$<$0.01\\
W$_{37,400}$ & & 1.07$\pm$1.18 & 0.33$\pm$0.11 & & 0.76$\pm$1.30 & 0.14$\pm$0.09 & & 2.18$\pm$4.29 & 0.32$\pm$0.17 & & 0.52$\pm$1.94 &$<$0.01\\
Way & & 0.94$\pm$0.90 & 0.31$\pm$0.11 & & 0.64$\pm$0.94 & 0.13$\pm$0.10 & & 1.88$\pm$3.22 & 0.30$\pm$0.17 & & 0.48$\pm$1.57 &$<$0.01\\
Ws & & 1.17$\pm$1.08 & 0.26$\pm$0.09 & & 0.82$\pm$1.14 & 0.10$\pm$0.07 & & 2.32$\pm$3.34 & 0.24$\pm$0.13 & & 0.46$\pm$1.03 &$<$0.01\\
NW & & 0.57$\pm$0.50 & 0.24$\pm$0.08 & & 0.38$\pm$0.53 & 0.11$\pm$0.08 & & 1.06$\pm$1.84 & 0.24$\pm$0.13 & & 0.26$\pm$0.94 &$<$0.01\\
CW & & 0.84$\pm$0.71 & 0.28$\pm$0.10 & & 0.55$\pm$0.73 & 0.12$\pm$0.09 & & 1.59$\pm$2.53 & 0.28$\pm$0.16 & & 0.39$\pm$1.27 &$<$0.01\\
BH & & 0.22$\pm$0.12 & 0.10$\pm$0.03 & & 0.11$\pm$0.13 & 0.04$\pm$0.03 & & 0.31$\pm$0.40 & 0.09$\pm$0.05 & & 0.06$\pm$0.16 &$<$0.01\\
\hline
\hline
\label{tab:sb}
\end{tabular}
\end{center}
\end{table*}

The main impact of assuming different IMFs (Way, Ws, central panel) is the fact that a top-heavy 
one, like the one from \cite{arimoto87}, produces much more SN-II, and to a lesser extent also more SN-Ia. This increases the 
abundance of all the metals and particularly O which is mainly ejected by SN-II \citep[we address the 
reader to the discussion in][for more details]{tornatore07}. This explains our results for the WHIM: since O has the 
peak of its emission lines at $T\approx10^6$ K and energies around 0.5 keV, the largest impact is seen in the 0.3--0.8 keV band 
surface brightness, where the WHIM is 20 per cent brighter in the Way run with respect to the Ws 
one (see also our results on line-emission in Sect.~\ref{sec:oxygen}). On the other side, we observe an opposite trend in the 
surface brightness of the IGM, with the top-light \cite{salpeter55} IMF having the highest emission. This result is due to the fact 
that when considering a top-light IMF less metals are produced, thus decreasing the cooling efficiency, and providing a higher 
fraction of hot $T > 10^7$ K gas that dominates the emission.

The most important variations are introduced when considering the different feedback schemes 
(bottom panel). 
Galactic winds have a twofold effect: on one side they suppress the peak emission 
in the central regions of clusters and, at the same time, they populate their outer parts with 
more material. Part of this gas is ejected at early epochs before the clusters form, while the 
remaining fraction is ejected directly in the intra-cluster medium significantly increasing the 
overall emission (see 
Fig.~\ref{fig:maps_tot}). As a result the IGM surface brightness is increased by a factor of 
$\sim2$ at all energies when comparing the NW and W runs, with the CW run constituting an 
intermediate scenario. The same applies to the WHIM emission but with only a $\sim20$ per cent 
increase from the NW to the W run. This is particularly evident at $E<0.8$ keV where oxygen 
emission dominates, indicating that winds are able to spread metals in the regions at intermediate 
temperatures.

On the other side, BH feedback greatly suppresses the emission, for both IGM and WHIM. In the 
IGM the difference range from a factor of 5 at lower energies up to one order of 
magnitude in the 2--10 keV band, while for the WHIM it is a factor of $\sim3$. This is the 
result of the strong BH winds that prevent gas from collapsing and suppress star formation.

We also estimate the cosmic variance associated to our results by computing the standard 
deviation (quoted as error in Table~\ref{tab:sb}) in 80 fields of 225 arcmin$^2$. For the 
IGM the cosmic variance is very high, comparable to the value itself: this happens because the 
presence of clusters and groups in the field dominates the total signal. Since the WHIM is 
dominated by the diffuse component its relative cosmic variance is lower, being of the order of 30 
per cent in the 0.3--0.8 keV band, but still higher than the expected differences introduced by 
the various physical prescriptions.

We can compare our results for the IGM\footnote{We do not compare our results on the WHIM given the 
different definitions (with and without the density cut at $\delta < 1000$) adopted in the two 
works.} also with our previous work \citep{roncarelli06}. Even if the feedback model of our W 
run is similar to the one adopted in that work \citep[see][]{borgani04}, their total expected 
emission in the 0.5--2 and 2--10 keV bands is higher by a factor of 2 and 3, respectively. This 
difference is due to the larger volume of the simulation that allowed to better sample the 
high-mass tail of the mass distribution, thus obtaining brighter haloes that dominate the emission.

When comparing our WHIM results with the ones of \cite{ursino10} in the 0.65--1 keV band, our 
predictions are close to their $0.15 \times 10^{-12}$ \sbunits\ obtained with a semianalytical 
characterization of the metallicity and a factor of $\sim 2$ lower than their higher 
metallicity models. Even if these results can be considered to be in a broad agreement, their 
differences reflect the current uncertainties in how metals spread in the lower density regions and 
on the ability of hydrodynamical models to provide precise predictions on the WHIM emission. 
More precisely, the typical metal abundances associated to our WHIM phase are of the order of 
$Z \approx 0.05$ \zsun, that corresponds to a regime where the expected signal is roughly equally 
distributed between free-free and line-emission. When assuming a higher metallicity ($Z=0.1-0.15$ 
\zsun) the emission becomes metal-dominated and increases significantly, thus justifying the 
differences in the results with the higher metallicity models of \cite{ursino10}.


\subsection{Comparison with the unresolved X-ray background} \label{sub:xrb}
Our results can be compared with the upper limits provided by the 
unresolved X-ray background (UXRB) from \cite{hickox07}, who measured the diffuse emission in the 
CDFs after subtracting all possible contaminating sources. However, we must consider that a 
direct comparison between our IGM maps/spectra and the measured UXRB is not possible since our 
mock maps include also clusters and groups that have been identified and eliminated in the CDFs 
observations. On the other side, using our WHIM maps can lead to misleading results since our WHIM 
definition is based on the intrinsic thermodynamical properties of the SPH particles, that do not 
correspond to an observational criterium of source identification \citep[see also the discussion 
in][]{roncarelli06}.

\begin{figure}
\includegraphics[width=0.475\textwidth]{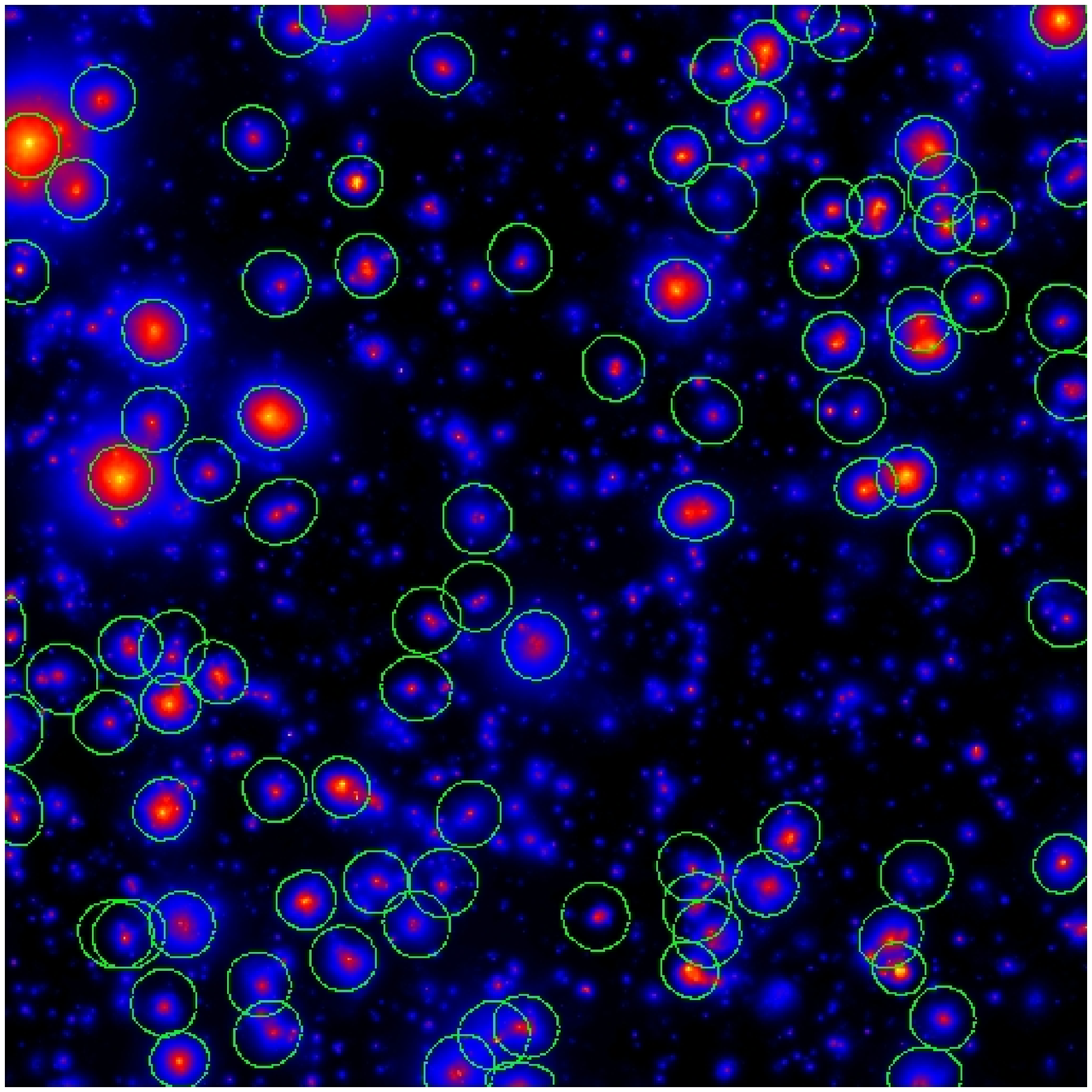}
\caption{Sketch of the haloes (marked by the elliptical regions) identified in one of our mock 
maps. The map represents the surface brightness (on a logarithmic scale) at 0.3 keV and 
corresponds to the same light cone shown in the maps of Figs.~\ref{fig:maps_tot} and 
\ref{fig:maps_dif}.}
\label{fig:mapreg}
\end{figure}

For these reasons we decide to adopt an observationally oriented approach to estimate the UXRB that 
corresponds to our simulations. Starting from our IGM maps/spectra (e.g., Fig.~\ref{fig:maps_tot}) 
we compute the photon counts in the 0.5--2 keV band expected from a \chandra\ Deep Field South 
(CDF-S) observation, by 
considering the ACIS-I response function at different energies (sampled in bins of 50 eV each).
In order to simulate reliable \chandra\ images we added an isotropic level of background that takes 
into account the contribution of both galactic and particle background. We neglected the 
contribution of undetected AGN/galaxies since, at flux limits of the CDF-S, their flux is completely 
described by point-like emission
associated with optical or IR-detected galaxies, which were excluded
by \cite{hickox07}. The images were then corrected for an average 
exposure of 4 Ms with an artificially added Poisson noise. The source identification has been  
performed by running a simple sliding cell detection with a cell size of 12 pixels and fixing a 
signal-to-noise ratio threshold of 4. We have then excluded all the regions within which the 
overall encircled signal from sources is above 4$\sigma$ with respect to the background. An example  
of the implementation of this method is shown in Fig.~\ref{fig:mapreg}.

\begin{figure}
\includegraphics[width=0.3\textwidth,angle=-90.]{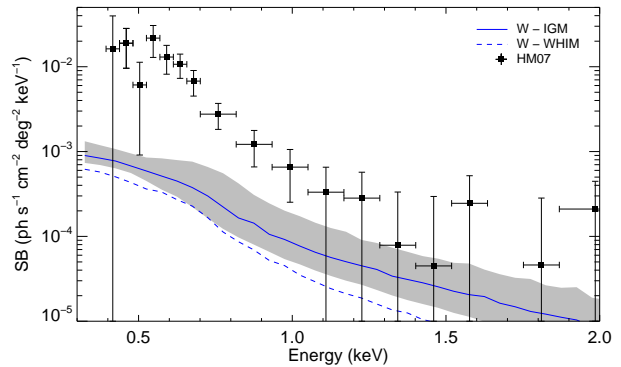}
\caption{Expected spectrum of the diffuse XRB associated to extragalactic gas emission for the W 
model. The solid line corresponds to the whole (IGM) gas emission, the dashed line to the WHIM 
contribution. The lines correspond to the median relative to 80 different values computed on an 
area of 225 arcmin$^2$ (e.g. CDFs), while the shaded region encloses the first and third 
quartile for the IGM. The black squares represent the data from \protect \cite{hickox07} with 
corresponding errorbars.}
\label{fig:specmask}
\end{figure}

Once we have identified all the pixels corresponding to detected haloes, we compute the expected 
spectrum of the remaining regions for all the different light-cones and simulations. We show in 
Fig.~\ref{fig:specmask} our results (averaged over the 20 light cones) for the W model for both the 
IGM and the WHIM and compare them with the observational estimates of \cite{hickox07}. The same 
comparison is shown in Table~\ref{tab:sb_mask} for the emission in three different X-ray bands and 
for the other physical prescriptions.

\begin{table*}
\begin{center}
\caption{The same as Table~\ref{tab:sb}, but after subtracting clusters and groups identified in a 
mock CDF-S observation. The last line indicates the observational measurements of \protect 
\cite{hickox07} with relative errorbars.}
\begin{tabular}{lccccccccc}
\hline
\hline
      & & \multicolumn{8}{c}{Surface brightness (10$^{-12}$ \sbunits)} \\
      & & IGM & WHIM & & IGM & WHIM & & IGM & WHIM  \\
Model & & \multicolumn{2}{c}{(0.5--2) keV} & & \multicolumn{2}{c}{(0.65--1) keV} & & \multicolumn{2}{c}{(1--2) keV}  \\

\hline
W       & & 0.57$\pm$1.31 & 0.17$\pm$0.21 & & 0.23$\pm$0.54 & 0.08$\pm$0.12 & & 0.23$\pm$0.67 & 0.04$\pm$0.06 \\
Way     & & 0.45$\pm$0.77 & 0.19$\pm$0.24 & & 0.18$\pm$0.34 & 0.09$\pm$0.13 & & 0.17$\pm$0.37 & 0.04$\pm$0.06 \\
Ws      & & 0.48$\pm$1.05 & 0.14$\pm$0.18 & & 0.20$\pm$0.49 & 0.06$\pm$0.09 & & 0.18$\pm$0.46 & 0.04$\pm$0.06 \\
NW      & & 0.33$\pm$0.42 & 0.16$\pm$0.17 & & 0.14$\pm$0.19 & 0.08$\pm$0.10 & & 0.11$\pm$0.20 & 0.03$\pm$0.04 \\
CW      & & 0.41$\pm$0.69 & 0.17$\pm$0.22 & & 0.17$\pm$0.31 & 0.08$\pm$0.13 & & 0.14$\pm$0.32 & 0.04$\pm$0.06 \\
BH      & & 0.20$\pm$0.18 & 0.08$\pm$0.05 & & 0.08$\pm$0.07 & 0.04$\pm$0.03 & & 0.07$\pm$0.09 & 0.02$\pm$0.01 \\
\hline
Obs.    & & \multicolumn{2}{c}{4.3$\pm$0.7} & & \multicolumn{2}{c}{1.0$\pm$0.2} & & \multicolumn{2}{c}{0.34$\pm$0.14}                \\
\hline
\hline
\label{tab:sb_mask}
\end{tabular}
\end{center}
\end{table*}

When looking at the results for our reference model, the surface brightness per unit energy of the 
UXRB associated to the IGM is well represented by a power-law (SB$\propto E^{-\gamma}$) with 
spectral index $\gamma=$0.9 (1.5, for the WHIM) in the 
0.3--0.8 keV band, with a clear steepening at higher energies, up to $\gamma=$2.5 (3.8) in the 
0.8--2 keV band. These values do not change significantly when other models are considered.

Through all the energy range our IGM predictions fall below the upper limits of the UXRB. At 
energies $E\lesssim 0.7$ keV the expected emission of the whole IGM is more than an order of 
magnitude below the observed UXRB. This indicates that at soft energies other sources are the main 
contributors, 
likely the emission associated to local components (Galaxy and Solar System). For these reasons the 
search for LSS emission in the UXRB at these energies requires a subtraction of these foreground 
components with more than 10 per cent accuracy. A possible observational technique consists in 
relying on their different angular correlation of the two components, as we will discuss in 
Section~\ref{sub:acf}.

On the contrary, we predict that at energies $E \gtrsim 1$ keV a significant fraction (from 50 per 
cent to all) of the UXRB is due to the emission of the IGM. The high value of the cosmic variance 
(reported as error in Table~\ref{tab:sb_mask}) together with the low angular surface of the 
CDFs does not allow us to provide more precise estimates.

When comparing our models with the different feedback and IMFs assumptions, the same 
considerations apply with respect to the discussion on Table~\ref{tab:sb}: 
galactic winds increase consistently the expected IGM signal.

It is important to point out that in all cases the surface brightness associated to a diffuse WHIM 
component is less than half of the total IGM one, ranging from $\sim$20 per cent to $\sim$40 per 
cent from high to low energies, with the only exception of the 50 per cent of the Way model in the 
0.65--1 keV band. This suggests that a consistent part of this signal is not associated with truly 
diffuse ($\delta < 1000$) gas. 
In particular this becomes important at $E \gtrsim 1$ keV where the majority of the UXRB due to the 
LSS is clearly associated to cluster outskirts and faint galaxy groups. However we must consider 
that LSS filaments also contain dense clumps that are 
excluded by our WHIM definition so, as a bottom line, we can consider our two estimates, IGM and 
WHIM, as bracketing the emission associated to the missing baryons component.


\subsection{The angular correlation function} \label{sub:acf}
Given the relatively low surface brightness of the WHIM, one of the possible techniques to 
disentangle its 
signal from the other brighter components (e.g. Galactic emission) consists in performing a 
correlation function analysis. To this purpose we compute the angular (auto)correlation function 
(\acf) $w(\theta)$ of the surface brightness in the 0.3--0.8 keV band.

We proceed as follows. First, for a given map we compute its surface brightness contrast
\begin{equation}
\delta(\vec{x}) \equiv (S_{\rm b}(\vec{x})/\overline{S_{\rm b}})-1 \ ,
\end{equation}
where $S_{\rm b}(\vec{x})$ is the surface brightness at the position $\vec{x}$ and 
$\overline{S_{\rm b}}$ is 
the average of the map. Then we compute the angular correlation function as
\begin{equation}
w(\theta)=\langle \delta(\vec{x}) \delta(\vec{x}+\vec{\theta})\rangle \ .
\end{equation}
With this definition a value of $w(\theta)=0$ corresponds to a random uncorrelated field.

\begin{figure}
\hspace{-0.2cm}\includegraphics[width=0.31\textwidth,angle=90]{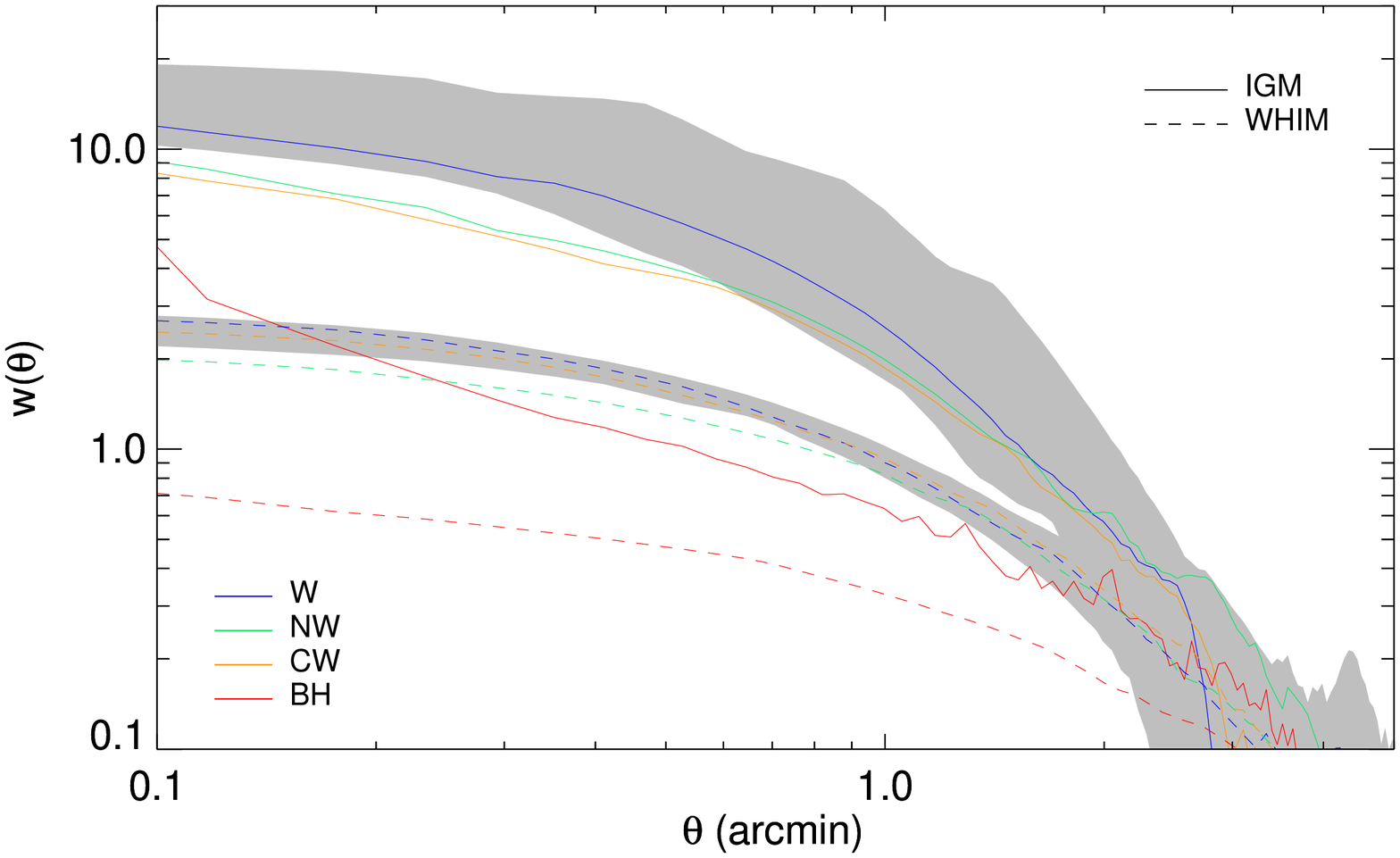}
\caption{Angular correlation function of the IGM (solid lines) and the WHIM (dashed) for different 
feedback models: the reference W model (blue line), the NW model (green), the CW model (orange) 
and the BH one (red). Each curve represents the median computed over the 20 independent 
light-cones (area of 900 arcmin$^2$) while the shaded regions enclose the first and third 
quartile.}
\label{fig:acf}
\end{figure}

We show in Fig.~\ref{fig:acf} our results for the IGM and the WHIM and for our 4 different feedback 
schemes.
As expected \citep[see, e.g.,][]{roncarelli06,ursino11}, the IGM correlates significantly up to 
angular scales of 2--3 arcmin, which correspond to the typical size of galaxy clusters and groups. 
At larger scales it drops rapidly to values consistent with zero ($\theta \gtrsim 5$ arcmin). 
Again, the cosmic variance produces significant deviations (grey-shaded area) from map to map. 
Galactic 
winds enhance the correlation signal of the IGM due to the higher relative clusters emission. The BH 
model produces a high correlation only at very low angular scales ($\theta \lesssim 0.2$ arcmin), 
consistently with what appears 
from Fig.~\ref{fig:maps_tot}. On the other side, at $\theta \gtrsim$ 2-3 arcmin the \acf\ is 
shallower due to the presence of a diffuse emission associated to gas ejected from galactic haloes at 
early epochs. When neglecting galactic winds (NW) high correlations at $\theta \gtrsim$ 2-3 arcmin are 
found as well. This is associated to the presence of small bright clumps that enhance the halo-halo 
correlation, and that show fainter peaks when winds are included (see the maps of 
Fig.~\ref{fig:maps_tot}).
The WHIM \acf\ is lower by almost one order of magnitude at low 
angular scales but has a shallower slope making it comparable to the IGM one for $\theta \sim 2$ 
arcmin. In this case, the different wind implementations (W, NW, CW) produce small variations 
in the \acf, of the order of 10 per cent.

We can compare our results with \cite{galeazzi09} who detected a putative WHIM emission 
by analysing the \acf\ of 6 different \xmm\ fields in the 0.4--0.6 keV band\footnote{We checked 
that the changes in the \acf\ associated to the two different bands are negligible.}. 
In this comparison one must take into account the different normalization adopted with respect to 
our work: in fact, since they do not know \emph{a priori} how much emission is associated to a 
diffuse component, they normalize $w(\theta)$ to the total UXRB flux, including also 
the foreground contribution. We can obtain an 
order-of-magnitude estimate of the impact of this difference by using our results on the 
average 0.4--0.6 keV surface brightness associated to unresolved IGM emission (see 
Section~\ref{sub:xrb}) and compare it to their total UXRB from \cite{hickox07}: the ratio between 
the latter and the former is $\sim 26$, which translates into a difference in the normalization of 
the order of $\sim 700$. If we apply this correction to the data of \cite{galeazzi09}, we obtain a 
value of $w(\theta) \simeq 7$ for $\theta$=1--2 arcmin, that falls within a factor of $\sim3$ 
from our IGM results. Given the high values of relative errors and cosmic variance associated to 
these measurements, we can not 
obtain a more precise estimate, but we can conclude that our results appear in broad agreement with 
those by \cite{galeazzi09} at small angular scales. On the other side, we obtain a steeper 
slope of the \acf\ at large angular scales, suggesting the possible presence of a 
contribution from other components in their data.


\section{Oxygen line emission}\label{sec:oxygen}

In this Section we focus on the properties of oxygen emission. In particular, we study 
the expected surface brightness due to O\vii\ and O\viii\ that represent the ions with the most 
significant contribution at X-ray energies.

\begin{figure}
\hspace{0.5cm}\includegraphics[width=0.30\textwidth,angle=-90.]{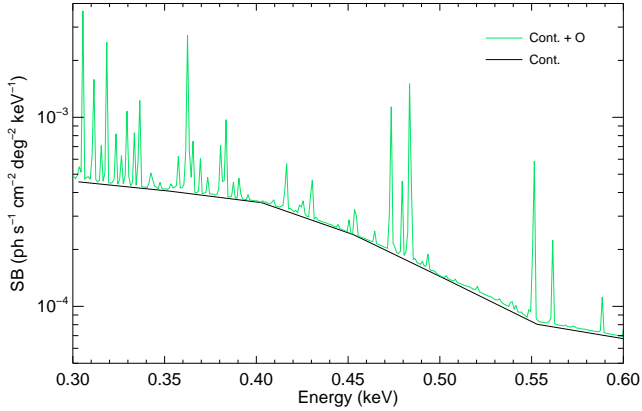}
\vspace{0.2cm}
\caption{Example of a spectrum extracted from a pixel of 1 arcmin$^2$ size, for the WHIM 
component only. The black line 
shows the continuum emission (i.e. including all the elements except oxygen) computed with 50 eV 
energy resolution. The green line represents the spectrum with the addition of oxygen emission, 
computed with 1 eV energy resolution to allow the identification of the lines.}
\label{fig:specO}
\end{figure}

To this purpose we take advantage of our method for computing the spectra associated to the SPH 
particles described by eq.~(\ref{eq:spec}) that allows us to separate the emission from the 
different elements. As discussed in Section~\ref{sub:coneset}, we use the last two light-cones set 
listed in Table~\ref{tab:cones}, which were computed with an energy resolution of 1~eV necessary 
for the analysis of single line emission. The energy range considered here (0.3--1 keV) allows us to 
enclose all of the most important O\vii\ and O\viii\ lines up to $z\simeq 0.9$, with no significant 
contribution from O\vi. In order to save memory, these O emission spectra have been computed with 
a poorer angular resolution of $\sim$1 arcmin, but this does not introduce any systematics since 
the expected correlation of the IGM and 
WHIM signal does not rise significantly at lower angular scales (see Section~\ref{sub:acf}). 
Moreover the expected angular resolution of the next-generation spectrographs (e.g., the 
\emph{IXO/ATHENA} proposal\footnote{
http://sci.esa.int/science-e/www/object/index.cfm?fobjectid=48729}) is higher than this limit. An 
example of a WHIM spectrum computed with our method for both oxygen-line and continuum emission is 
shown in  Fig.~\ref{fig:specO}.

We compute the statistics of the expected number of O lines (O\vii\ and O\viii\ added together) 
above a given surface brightness along the line of sight. We proceed in the 
following way. For a given pixel of our maps, that corresponds to a line-of-sight of 0.88 
arcmin$^2$ and extending in the range $0<z<1.5$, we consider its spectrum in the energy interval 
0.3--1 keV and identify the 1~eV bins corresponding to a surface 
brightness higher than a threshold fixed to SB$_{\rm th}=2 \times 10^{-6}$ \sbphot\ (safely below 
the detection limits of future instruments). Then, since an emission line can extend to more than 
1~eV, we assign to a single line all the contiguous bins that fall above this threshold and sum 
their values to obtain the total surface brightness of the line itself. We repeat 
the same procedure for all the pixels and all the 20 different light-cones for both the IGM and the 
WHIM. We checked that the final results do not change significantly when degrading our angular 
resolution to $\sim$2~arcmin or when changing the value of SB$_{\rm th}$.

\begin{figure}
\includegraphics[width=0.30\textwidth,angle=-90.]{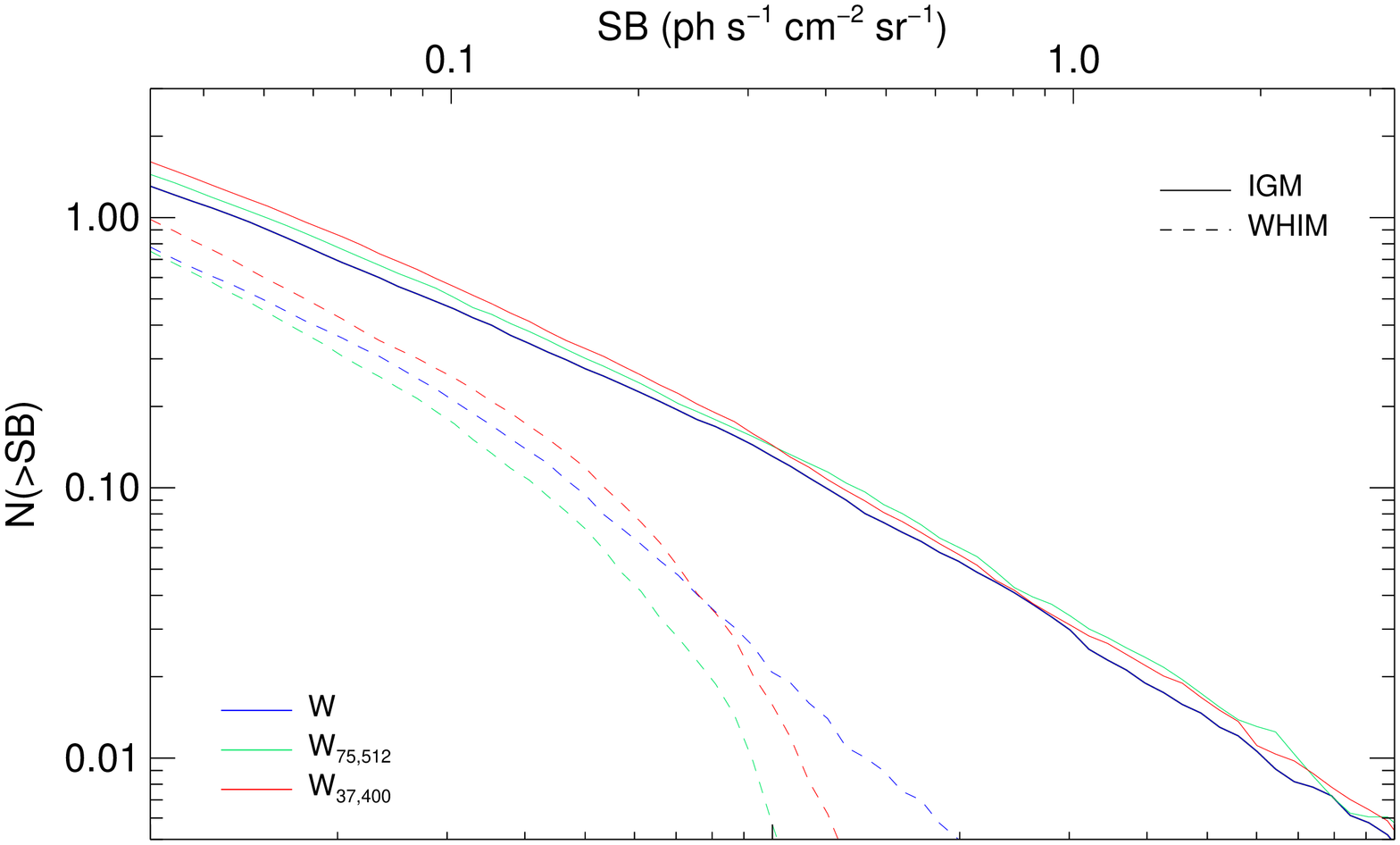}
\includegraphics[width=0.30\textwidth,angle=-90.]{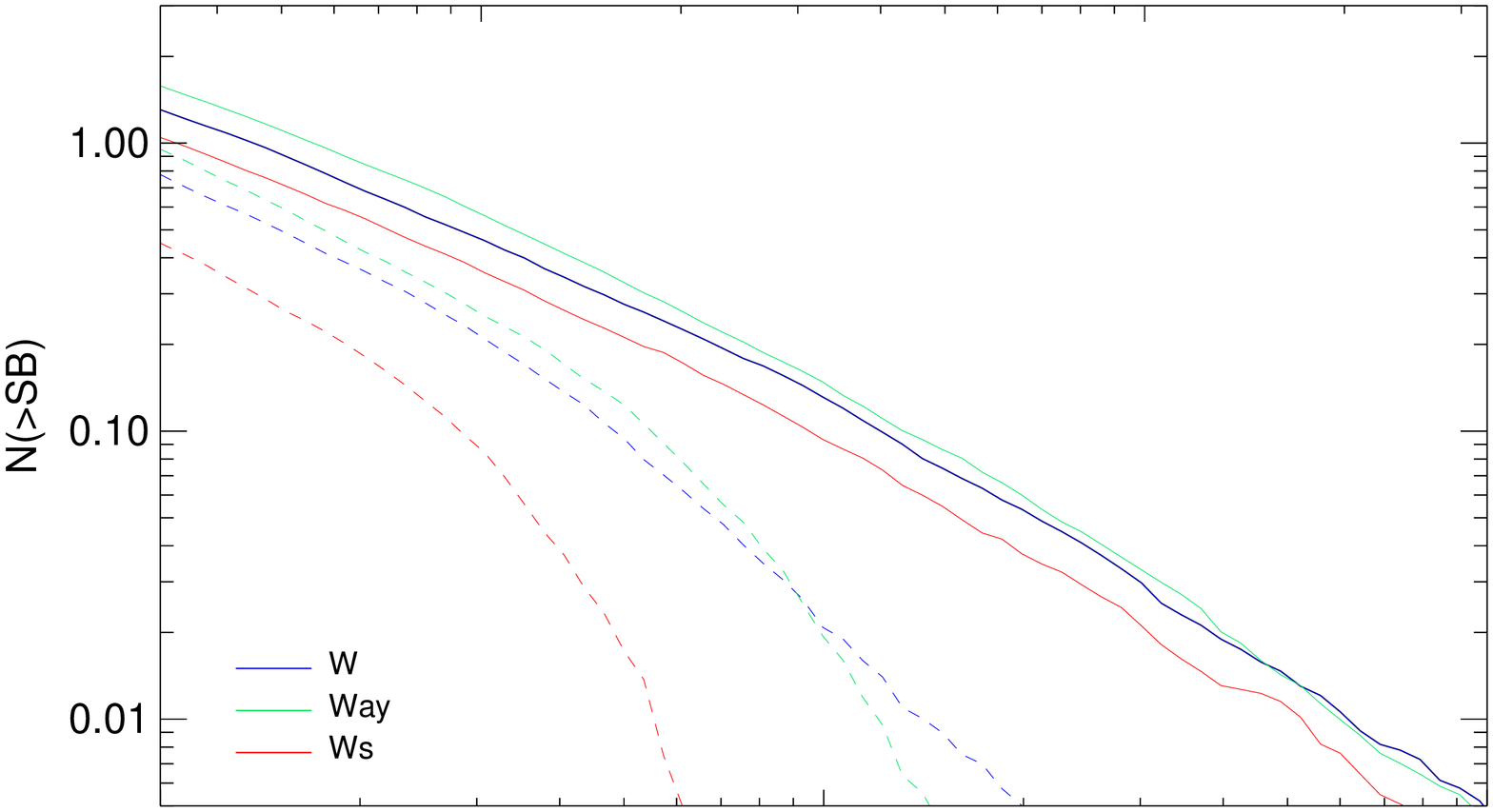}
\includegraphics[width=0.30\textwidth,angle=-90.]{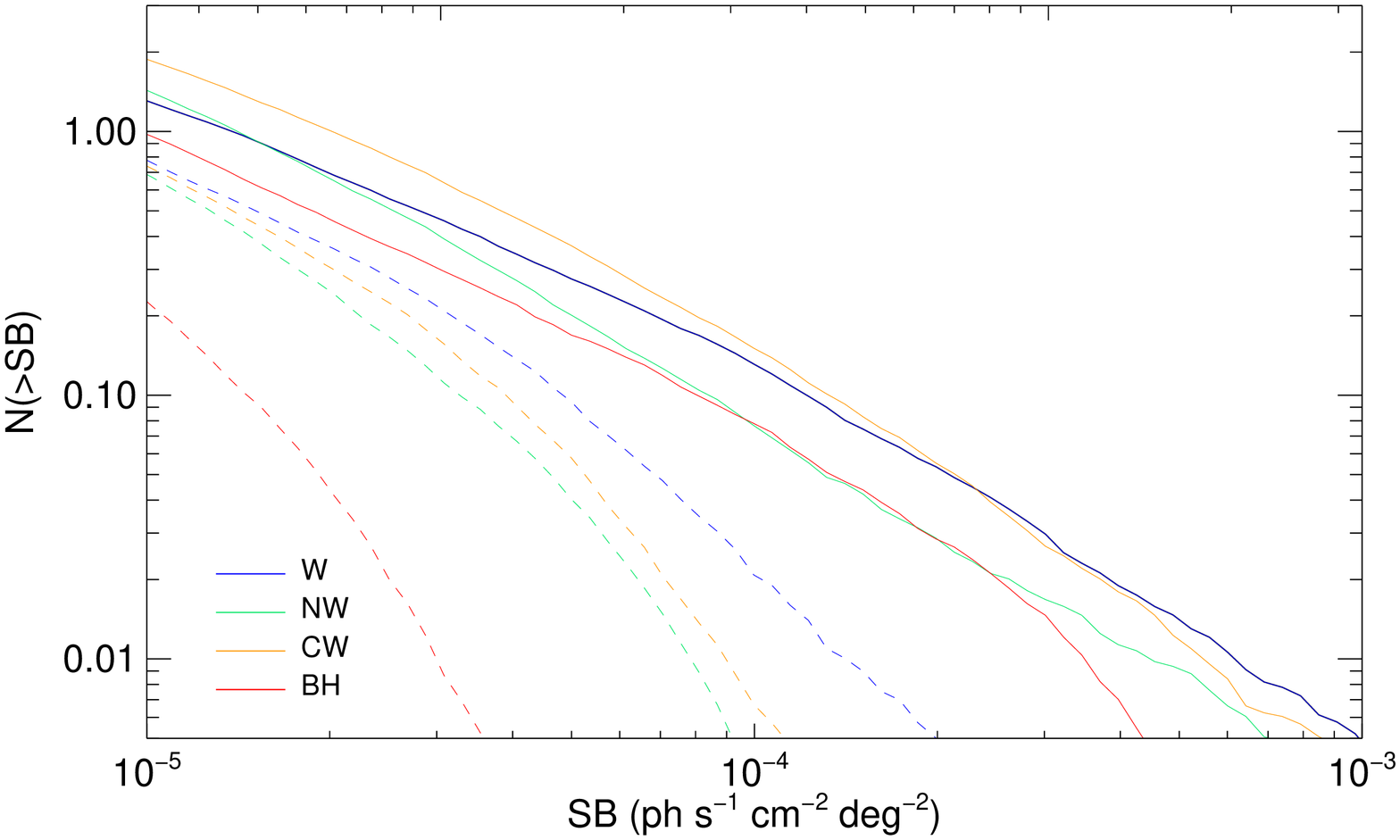}
\caption{Expected counts of oxygen (O\vii\ and O\viii) line emission associated to the IGM (solid 
lines) and WHIM (dashed lines) as a function of lines brightness. In all plots the blue 
lines correspond to our reference model (W). Top panel: comparison with \runone\ (green) and 
\runtwo\ (red). Central panel: comparison with Way (green) and Ws (red). 
Bottom panel: comparison with NW (green), CW (orange) and BH (red).}
\label{fig:olines}
\end{figure}

Fig.~\ref{fig:olines} shows the expected number counts of oxygen lines for our reference run 
compared with the other ones.
When considering the different simulation parameters (top panel) we see that increasing 
the mass resolution of the simulation by a factor of $\sim$4 enhances the expected line counts for 
the IGM by $\sim$20 per cent. This is due to improved capability of resolving the high density 
regions that host the star-forming metal-rich gas and are therefore bright line-emitters. The 
effect is also present when considering the WHIM oxygen lines in the faint (SB $< 10^{-4}$ \sbphot) 
end. This suggests that a high resolution is mandatory in this kind of analysis and is more 
desirable than a large simulation volume.

The different IMFs (central panel) produce changes in the amount of SN-II and, consequently, on the 
total amount of oxygen released by stars. With the top-heavy \cite{arimoto87} IMF 
(Way model) the increased amount of SN-II produces significantly more oxygen. If we take as a 
reference a surface brightness detection threshold of 3$\times 10^{-5}$ \sbphot\ (a possible goal 
of next-generation instruments), this translates into $\sim$20 per cent more detected lines for 
both IGM and WHIM. The opposite effect happens when adopting the \cite{salpeter55} IMF (Ws).

The changes introduced by different feedback schemes (bottom panel) are also relevant. 
When suppressing galactic winds (NW) metals remain more concentrated in the cold star-forming 
regions where they are not able to emit significantly: this results in a factor of 2 less 
expected lines for the IGM. The CW model shows trend for the IGM similar to the W model, with a 
small  increase for fainter lines.
When looking at the WHIM curves, the effect of metal transport due 
to galactic winds can be seen with the higher amount of lines expected with respect to the NW 
model, that reaches an order of magnitude for brighter (SB$> 10^{-4}$ \sbphot) lines.
Coupled winds are not able to diffuse efficiently metals in the low density regions, resulting in a 
number of lines closer to the NW run than to the W one. Considering the BH feedback, the expected 
line counts are significantly 
smaller for both IGM and WHIM. Again, this is the result of the early energy feedback that 
suppresses the gas collapse resulting in the IGM having less dense structures. This effect is even 
more remarkable in the WHIM.


\subsection{Comparison with other models}\label{sub:comp_o}
Apart from the differences introduced by the various feedback mechanisms, we can compare the 
results of our reference run with the expected oxygen line counts estimated by \cite{takei11} and 
\cite{cen06b}. On the whole, the number of oxygen lines that we estimate is significantly lower 
with respect to their results. If we consider the B1 model of \cite{takei11} and a reference 
surface brightness of 0.1 \sbphotsr, they obtain a value of $N \simeq 2$ for the sum 
of O\vii\ and O\viii\ lines in the IGM case and $N \simeq 1$ for the WHIM\footnote{The authors 
show the value of $\Delta N / \Delta z$. Since their redshift interval is $0<z<0.5$ here we 
consider that the contribution from lines associated to higher redshift emitters is 
negligible.}, while our reference model predicts 0.5 and 0.25 respectively, a factor 3--4 lower. 
As a consequence, with our results the amount of WHIM emission lines detectable by future 
X-ray instruments is significantly reduced. If we consider a resolution element of 2.6 arcmin, our W 
model predicts $\sim$100 detections per square degree of either O\vii\ or O\viii\ lines from the WHIM.
The reason of this discrepancy has to be searched in the different recipe adopted to describe the 
metal enrichment of the gas. \cite{takei11} adopt a post-processing method assigning a metallicity 
to the SPH particles as a function of overdensity: in particular, they use the results 
of \cite{cen99} who estimated a relation of the form $Z \propto \delta^{1/2}$. If we compare it to 
what was shown in \cite{tornatore10} we can see that this results in a factor of $\sim2$ less metals at 
overdensities $\delta < 10^4$. In addition to that, if we take into account also the differences in 
the relative element abundance, our model presents a O/Fe mass ratio of $\sim 2.5$ while 
\cite{takei11} adopt solar yields from \cite{anders89} which correspond to a mass ratio of 5.2. 
This adds another difference by a factor of 2 in the abundance of oxygen. At a first approximation, 
given the linear dependence of line emissivity on the element abundance, this translates into a 
factor of $\sim$4 in line surface brightness that is sufficient to explain the differences 
with their work. In order to verify this hypothesis we run a test light-cone implementing their 
$Z-\delta$ relation and, as expected, we obtained results similar to \cite{takei11}.

We can also compare our results with \cite{cen06b} who computed the expected oxygen lines by using 
a hydrodynamical simulation. Even if they provide only the value of $dN/dz$ at $z=0$, we can 
estimate that our numbers are smaller, coherently to what said about \cite{takei11}. Again, part of 
the difference is due to the different element yields: \cite{cen06b} artificially augment the 
oxygen abundance of their model (see the discussion in their Section 2) to reach a value of O/Fe 
close to 2 times solar\footnote{We converted their ratio of 1.67 referred to \cite{anders89} yields 
to the ones of \cite{asplund09}.}. Such a high value is ruled out by current estimates from observed galaxy 
clusters \citep{tamura04,simionescu09}. In addition to that, their simulation is able to spread 
more efficiently the metals outside the hot and dense regions, resulting in more oxygen in the 
thermodynamical conditions that correspond to bright O\vii\ and O\viii\ emission. 
This issue is connected to the current uncertainties in the hydrodynamical codes and particularly 
in the feedback implementation. More in detail, it has been shown that introducing a coupled winds 
scheme on simulations 
at galactic scales can impact galactic properties as well as the amount of ejected material 
\citep{dallavecchia08}. Moreover also the implementation of momentum-driven winds in hydrodynamical 
simulations can modify the star-formation history and increase the amount of metals ejected at earlier 
epochs \citep[see, e.g.,][]{oppenheimer06,tescari09}. However, given the complex interplay between wind 
efficiency and star formation regulation, it is difficult to determine whether these models might reduce 
the discrepancy between our results and the ones of \cite{cen06b}.
In this framework, an interesting comparison can be provided by estimates from O\vi\ absorption systems 
that are, however, beyond the scope of this paper.


\section{Conclusions}\label{sec:concl}
In this paper we have analyzed a set of cosmological hydrodynamical simulations presented in  
\cite{tornatore10} that include a detailed treatment of chemical enrichment of the IGM. The aim is 
to provide a description of the expected X-ray emission of the WHIM, with focus on the influence of 
different feedback mechanisms, namely initial mass functions, galactic winds and black hole 
feedback.

Our results can be summarized as follows.
\begin{itemize}
\item[(i)]
Galactic winds have a strong impact on the expected signal of the WHIM and the IGM in general: with 
our assumptions ($v_w=500$ km/s, $\eta=2$) the surface brightness is enhanced by a factor of 2 
through all the energy range of interest (0.3--10 keV). This is due to the ability of winds to push 
gas and metals in cluster outskirts and outside, making lower density regions brighter.
\item[(ii)]
On the other side BHs, acting at earlier epochs, can prevent the IGM to collapse in dense 
structures suppressing significantly the expected emission of clusters and WHIM.
\item[(iii)]
All of our models predict that the expected contribution from non-resolved LSS structures (galaxy 
groups, WHIM) is below the upper limits of the UXRB \citep{hickox07}. At energies 
$E\lesssim0.7$ 
keV they account for only $\sim$ 10 per cent of the UXRB, indicating that Galactic components 
dominate, while at $E \gtrsim 1$ keV, they account from half to all of the signal, with significant 
field-to-field differences expected due to cosmic variance.
\item[(iv)]
The WHIM shows an angular correlation function $w(\theta)>1$ up to scales of 2-3 arcmin, with 
very low variations due to feedback effects or cosmic variance. At low angular scales ($\theta 
= 1-2$ arcmin), our results are in broad agreement with the observations of \cite{galeazzi09}, 
while for larger angles we obtain lower values.
\item[(v)]
The expected oxygen (O\vii\ and O\viii) line counts depend significantly on the choice of the 
feedback models. In particular, the number of detectable line counts is increased by a factor of 3 
by GWs and by 20 per cent when assuming a top-heavy IMF \citep{arimoto87}. We obtain also a 
significant enhancement by increasing the mass resolution of the simulation.
\item[(vi)]
With our simulations we predict a number of detectable oxygen emission lines lower by a factor of 
3-4 with respect to the predictions of other works \citep{cen06b,takei11} due to  
differences in the implementation of chemical enrichment in the hydrodynamical 
codes. This has to be subject of further investigations and has to be considered a part of 
current uncertainties in the modelisation of the WHIM properties. However, if our results were 
confirmed, observing O lines from the WHIM might be a difficult task even for the upcoming X-ray 
observatories.
\end{itemize}

As a conclusion, our work confirms the importance of the use of hydrodynamical simulations in the 
modelisation of the LSS properties, together with the comparison with observations. In particular, 
we highlighted how the new generation of X-ray telescopes could provide constraints on both the 
missing baryon component and the mechanisms of star-formation and feedback that affect the 
WHIM. In this framework, it will be interesting to extend the results of our work also to the 
absorption properties in the X-ray and UV frequency range.


\section*{acknowledgments}
Most of the computations necessary for this work have been performed thanks to the Italian
SuperComputing Resource Allocation (ISCRA) of the {\it Consorzio Interuniversitario del Nord Est 
per il Calcolo Automatico} (CINECA). We acknowledge financial contributions from contracts 
ASI-INAF I/023/05/0, ASI-INAF I/088/06/0, ASI I/016/07/0 COFIS, ASI Euclid-DUNE I/064/08/0, ASIUni 
Bologna-Astronomy Dept. Euclid-NIS I/039/10/0, PRIN MIUR ``Dark energy and cosmology with large 
galaxy survey'', the European Commissions FP7 Marie Curie Initial Training Network CosmoComp 
(PITN-GA-2009-238356), by the PRIN-MIUR09 ``Tracing the growth of structures in the Universe'' and 
by the PD51 INFN grant. We thank an anonymous referee that provided useful comments which contributed 
to improve the presentation of our results. We acknowledge useful discussions with A.~Baldi, S.~Di~Meo, 
S.~Ettori, D.~Fabjan, R.~Hickox, L.~Lovisari, R.~Smith, L.~Tornatore and E.~Ursino. We are grateful to 
R.~Hickox for providing us the data on the UXRB. The parallelization of the light-cone simulator 
code has been done with the assistance of C.~Gheller. A special thank goes to D.~Fabjan for the 
help in dealing with simulation outputs.

\bibliographystyle{mn2e}

\label{lastpage}
\end{document}